\RequirePackage{fix-cm}
\documentclass[smallcondensed]{svjour3}     
\smartqed  

\usepackage{graphicx}
\usepackage{xcolor}
\usepackage{amssymb}
\usepackage{cite}
\usepackage{subcaption}
\usepackage{hyperref}
\usepackage{comment}

\begin{document}

\title{Comparative Evaluation of Community-aware Centrality Measures
}


\author{Stephany Rajeh        \and
        Marinette Savonnet \and Eric Leclercq \and Hocine Cherifi.
}


\institute{Laboratoire d’Informatique de Bourgogne - University of Burgundy, Dijon, France\\
             \email{stephany.rajeh@u-bourgogne.fr} }


\maketitle

\begin{abstract}
Influential nodes play a critical role in boosting or curbing spreading phenomena in complex networks. Numerous centrality measures have been proposed for identifying and ranking the nodes according to their importance.   Classical centrality measures rely on various local or global properties of the nodes.   They do not take into account the network community structure. Recently, a growing number of researches have shifted to community-aware centrality measures.  Indeed, it is a ubiquitous feature in a vast majority of real-world networks. In the literature, the focus is on designing community-aware centrality measures.  However, up to now, there is no systematic evaluation of their effectiveness.  This study fills this gap. It allows answering which community-aware centrality measure should be used in practical situations. We investigate seven influential community-aware centrality measures in an epidemic spreading process scenario using the Susceptible-Infected-Recovered (SIR) model on a set of fifteen real-world networks. Results show that generally, the correlation between community-aware centrality measures is low. Furthermore, in a multiple-spreader problem, when resources are available, targeting distant hubs using Modularity Vitality is more effective. However, with limited resources, diffusion expands better through bridges, especially in networks with a medium or strong community structure.
\end{abstract}

\section{Introduction}
Complex networks can describe a wide range of real-world complex systems such as power grids, air transportation, social networks, and financial transactions. Influential nodes identification within these systems is of great interest. Indeed, one can use this knowledge to enhance the diffusion process in viral marketing applications and control epidemic spreading through appropriate immunization strategies. Numerous centrality measures have been proposed to identify influential nodes \cite{lu2016vital}. Local centrality measures compute the centrality of a node based on its neighborhood. Global centrality measures quantify the centrality of a node by inspecting its position in the entire network. Finally, mixed centrality measures combine both types of information \cite{berahmand2019new}. These centrality measures, which we call classical, exploit various topological properties of the nodes ignoring the network organization in communities. However, it is well-known that community structure is one of the main features characterizing real-world networks \cite{girvan2002community}.

Current developments exploit the modular organization of networks to derive community-aware centrality measures \cite{ghalmane2019immunization, guimera2005functional, tulu2018identifying, gupta2016centrality, modvitality, zhao2015community, luo2016identifying}. Indeed, the structuring of networks in communities is ubiquitous in many real-world networks. Furthermore, it significantly affects the diffusion dynamics of a network \cite{cherifi2019community}.

Consequently, nodes that may not be considered influential by a classical centrality measure (i.e., agnostic about the community structure) may be of ultimate influence when one considers the mesoscopic organization of the network. Community-aware centrality measures distinguish intra-community links from inter-community links. The former links join nodes in the same community while the latter join nodes in different communities. Intra-community links allow quantifying the node's local influence inside its community. In contrast, inter-community links account for the node's global effect on the various communities forming the network.

Community-aware centrality measures differ on how they combine the two types of links. Community Hub-Bridge \cite{ghalmane2019immunization} picks up highly connected nodes (hubs) and bridges (connections between communities) simultaneously. It weights the intra-community links with the node's community size.  The weight of inter-community links is the number of neighboring communities. With a slightly different approach, Comm Centrality \cite{gupta2016centrality} gives more importance to bridges by raising the inter-community links to the power of 2. Community-based Centrality \cite{zhao2015community} ponderates the intra-community links and the inter-community links depending on their respective community sizes. For example, suppose a node has an inter-community link connecting to a community made up of more than 80\% of the total nodes in a network. This specific inter-community link will receive a larger weight than another inter-community link connecting to a smaller community. K-shell with Community \cite{luo2016identifying} identifies important nodes based on their $k$-shell value by considering intra-community links and inter-community links separately. Participation Coefficient puts more emphasis on the heterogeneity of inter-community links \cite{guimera2005functional}. If a node participates in several communities, it will receive a high participation coefficient. Similarly, Community-based Mediator \cite{tulu2018identifying} targets influential nodes based on the heterogeneity of their intra-community and inter-community links. However, in this case, it combines the links using entropy. If intra-community and inter-community links of a node are equal, Community-based Mediator reduces to the normalized degree centrality. Modularity Vitality \cite{modvitality} is based on the modularity measure of the community structure strength. It computes the node centrality using the modularity variation induced when one removes it from the network. Modularity Vitality distinguishes bridges from hubs. The idea is that eliminating hubs decreases modularity while removing bridges increases it.

In previous works \cite{rajeh2021characterizing, rajeh2021correlated}, we studied the interactions between classical and community-aware centrality measures and the influence of the network topological features. Results show that the community structure strength is the main feature driving the correlation between classical and community-aware centrality measures.

In contrast, this work focuses on the interactions between community-aware centrality measures. We perform a systematic investigation of seven popular community-aware centrality measures using fifteen real-world networks. These networks originate from various domains and cover a wide range of basic topological properties. First, we perform a correlation analysis of the multiple pairs of community-aware centrality measures. Second, we investigate the influence of the community structure strength. Finally, we analyze the effectiveness of the community-aware centrality measures. To this end, we consider their relative diffusive power in a scenario based on the Susceptible-Infected-Recovered (SIR) model using a multiple-spreader infection scheme. This extensive comparative analysis provides new insights to preliminary investigations conducted in online social networks in the SIR single-spreader infection scheme reported in \cite{rajeh2021comparing}.

The main contributions of the paper summarize as follows:
\begin{enumerate}
\item We investigate the correlation of several influential community-aware centrality measures.
\item We perform an extensive comparative analysis of their effectiveness in a SIR simulation scenario using real-world networks from different domains with a wide range of topological properties. 
\item We study the interplay between the community structure characteristics, the availability of resources, and the performance of the various community-aware centrality measure under test.
\item We give clear indications about the most effective strategies to employ according to the community structure strength and the availability of resources at hand.
\end{enumerate}


The paper is organized as follows. Section \ref{sec:RelatedWork} briefly discusses the related work. The community-aware centrality measures are introduced in section \ref{sec:CommAwareCent}. The data and the tools used in the evaluation process are presented in section \ref{sec:DataToolsAndMethods}. In section \ref{sec:ExpResults}, the experimental results are given. In section \ref{sec:Discussion}, a discussion on the findings is developed. Section \ref{sec:Conc} concludes the article.

\section{Related Work}
\label{sec:RelatedWork}
The relationship between centrality measures and topological network features is a fundamental issue. Initially focused on macroscopic topological properties, it has recently shifted to mesoscopic topological properties. One can categorize related works into three main topics. The most widespread research works deal with the interplay between classical centrality measures and networks' macroscopic topological features. Uncovering the relations between essential network features and community structure is also a well-studied issue. Finally, although there is a growing trend to design community-aware centrality measures, to our knowledge, there are few exhaustive comparative evaluations.

Studies that investigate the relationship between networks' macroscopic topological properties and centrality measures concern mainly classical centrality measures \cite{li2015correlation, ronqui2015analyzing, rajeh2020interplay, schoch2017correlations, oldham2019consistency}. In \cite{li2015correlation}, Li et al. study the correlations between the betweenness, the closeness, the components of the principal eigenvector, the degree, the first-order degree mass, and the second-order degree mass, in the Erd\~{o}s-R\'enyi and the scale-free network models and 34 real-world networks. They show that classical centrality measures are generally positively correlated independently of the network's size. Ronqui and Travesio \cite{ronqui2015analyzing} show that classical centralities have stronger correlations in scale-free network models such as the Barabási–Albert model. Rajeh et al. \cite{rajeh2020interplay}, observe that the higher the network's density and transitivity, the higher the correlation between centrality measures and hierarchy measures ($k$-core and $k$-truss). Schoch et al. \cite{schoch2017correlations} show that the closer a graph to a threshold graph, the more correlated classical centrality measures, independent of their conceptual distinctness. Note that a threshold graph is a class of graphs that generalizes the star graph property and represents the purest form of a core-periphery structure. Indeed, one can partition the nodes into a clique and an independent set. Oldham et al. \cite{oldham2019consistency} show that modularity plays a significant role in the correlation between classical centrality measures. Although these works give precious insights on how network topology affects the redundancy between centrality measures, they do not investigate community-aware centrality measures.

Uncovering the relations between the macroscopic topological properties and the community structure of networks has also attracted many researchers \cite{nematzadeh2014optimal, orman2013empirical, wang2010impact, wharrie2019micro,lancichinetti2010characterizing}. For example, Nematzadeh et al. \cite{nematzadeh2014optimal} show that a strong community structure enhances information diffusion under the linear threshold model through intra-community links. Orman et al. \cite{orman2013empirical} demonstrate the positive relationship between the community structure strength and transitivity. The work of Wang et al. \cite{wang2010impact} illustrates that communicability is more difficult in networks with a well-defined community structure. Wharrie et al. \cite{wharrie2019micro} show that the number of communities naturally increases when networks exhibit high clustering. Lancichinetti et al. \cite{lancichinetti2010characterizing} report that networks from the same domain often have several similar mesoscopic characteristics, such as the community size distribution. These studies demonstrate the interplay between the community structure and the macroscopic topological properties of the network. Nonetheless, they do not relate these outcomes with community-aware centrality measures.

Even though various community-aware centrality measures have been developed \cite{ghalmane2019immunization, guimera2005functional, tulu2018identifying, gupta2016centrality, modvitality, zhao2015community, luo2016identifying, rajeh2020investigating }, the literature reports few comparative studies. Authors generally introduce a new community-aware centrality measure and compare it with a few alternatives. Furthermore, there is no common framework to assess their performance. In recent works \cite{rajeh2021characterizing, rajeh2021correlated}, we performed extensive comparative studies to evaluate the relationship between classical and community-aware centrality measures. Results show that the community structure strength is the top network feature influencing their correlation. More specifically, in networks with a strong community structure, the local influence of a community-aware centrality measure highly correlates with its classical counterpart. In contrast, correlation is low for the global impact of the community-aware centrality measures. One observes the opposite behavior in networks with a weak community structure.

In preliminary work, we investigated the diffusion dynamics of seven community-aware centrality measures on online social networks. This study uses the Susceptible-Infected-Recovered (SIR) model under a single-spreader activation scheme \cite{rajeh2021comparing}. Results give precious indications about the effectiveness of each measure for identifying top spreaders. However, more experiments are necessary to consolidate these results. Here, we extend this work in three directions. First, we consider a multiple-spreader activation scheme for the diffusion process. Second, we use additional networks from various domains and different topologies and sizes. Third, we investigate the influence of the community structure strength in correlating the community-aware centrality measures.

\section{Community-aware Centrality Measures}
\label{sec:CommAwareCent}

The definitions of the seven community-aware centrality measures are provided in this section. Assume that $G(V,E)$ is a simple, undirected, and unweighted graph where $V$ is the set of nodes of size $N=|V|$ nodes and $E \subseteq V \times V$ is the set of edges of size $M=|E|$ edges. The graph $G$ is partitioned into $\varsigma=\{c_1, c_2, ... , c_q, ..., c_{|\varsigma|}\}$ communities, where $C=|\varsigma|$ is the total number of communities, $c_q$ is the $q$-th community, and $n_{c_q}$ is the number of nodes in community $c_q$. Each node $i$ in $G$ has a total degree of $k_i^{tot}$ = $k_i^{intra}$ + $k_i^{inter}$ where  $k_i^{intra}$ and  $k_i^{inter}$ are the intra-community and inter-community links, respectively. Additionally, $k_{i,c_q}$ is the number of links node $i$ has in a given community $c_q$. Table \ref{TableNotations} reports the list of symbols used in this paper and Table \ref{TableCharacteristcs} states the main characteristics of the community-aware centrality measures under investigation.

\textbf{1. Community Hub-Bridge} \cite{ghalmane2019immunization} targets hubs and bridges simultaneously. It weights the node's intra-community links by the size of its community and the inter-community links by the number of communities a node can reach in one hop. It is defined as follows:

\begin{equation}
\alpha_{CHB}(i) = n_{c_q} \times k_i^{intra} + \sum_{c_l \subset \varsigma \backslash  c_q}^{N} \bigvee_{j \in c_l} a_{ij}  \times k_i^{inter}
\end{equation}

where $n_{c_q}$ is node $i$'s community size and $\bigvee_{j \in c_l} a_{ij} = 1$ if node $i$ is connected to at least one node $j$ in community $c_l$.

\textbf{2. Participation Coefficient} \cite{guimera2005functional} emphasizes a node's influence based on the fraction of its inter-community links.  If a node is linked only to nodes in its community, its Participation Coefficient is zero.  It is defined as follows:
\begin{equation}
\alpha_{PC}(i) = 1 - \sum_{q=1}^{C} 
\left(
\frac{k_{i,c_q}}{k_i^{tot}}
\right)^2
\end{equation}

\textbf{3. Community‑based Mediator} \cite{tulu2018identifying} is based on the entropy of the intra-community and inter-community links of a node. The more mixed the links of a node, the higher its centrality value. It is defined as follows:
\begin{equation}
\alpha_{CBM}(i) = H_i \times \frac{k_i^{tot}}{\sum_{i=1}^{N} k^{tot}_i}
\end{equation}

where $H_i$=[$-\sum \rho_i^{intra} log(\rho_i^{intra})]$+$[- \sum \rho_i^{inter} log(\rho_i^{inter})$] is the entropy of node $i$ based on its $\rho^{intra}$ and $\rho^{inter}$ which represent the node's ratio of intra-community and inter-community links over the total degree and $\sum_{i=1}^{N} k^{tot}_i$ is the total sum of all degrees in the network.

\textbf{4. Comm Centrality} \cite{gupta2016centrality} preferentially targets bridges. However, hubs locally situated with their intra-community links are not discarded. It is defined as follows:
\begin{equation}
\alpha_{Comm}(i) =  (1 + \mu_{c_q}) \times \chi +  (1 - \mu_{c_q})  \times  \varphi^2
\end{equation}
where $\mu_{c_q}$ is the fraction of inter-community links over the total community links in community $c_q$, $\chi = \frac{k_i^{intra}}{max_{(j \in c)}k_j^{intra}} \times R$, $\varphi = \frac{k_i^{inter}}{max_{(j \in c)}k_j^{inter}} \times R$, and $R$ is a constant to standardize intra-community and inter-community values.

\textbf{5. Modularity Vitality} \cite{modvitality} is a signed community-aware centrality measure. It can differentiate a hub from a bridge based on Newman's modularity \cite{newman2006modularity} variation after the node's removal. It is defined as follows:
\begin{equation}
\alpha_{MV}(i) =  Q(G) - Q(G \setminus \{i\})
\end{equation}

where $Q(G)$ is the network's modularity and $Q(G \setminus \{i\})$ is the network's modularity after the removal of node $i$. Note that in this study, we use the version of Modularity Vitality which targets hubs first (i.e., nodes are ordered from positive to negative magnitude). We denote it as $\alpha^+_{MV}$. Results with other versions targeting bridges first are not reported because they show poor performance. Note that computation of the Modularity Vitality upon node removal involves limited information. Indeed, it only requires knowledge about the node's 1-hop neighborhood and the total degrees of communities.

\textbf{6. Community-based Centrality} \cite{zhao2015community}  weights the node's intra-community and inter-community links by the size of their communities. It is defined as follows:

\begin{equation}
\alpha_{CBC}(i) =  \sum_{q=1}^{C} k_{i,c_q} 
\left(
\frac{n_{c_q}}{N}
\right)
\end{equation}

\textbf{7. K-shell with Community}\cite{luo2016identifying} splits the network $G$ into two networks. One is made of the nodes and their intra-community links, and the other contains the nodes and the inter-community links. Then, a linear combination of the $k$-shell hierarchical decomposition of these networks assesses the node's influence \cite{dorogovtsev2006k}. It is defined as follows:

\begin{equation}
\alpha_{ks}(i) = \delta \times \alpha^{intra}(i) + (1- \delta) \times \alpha^{inter}(i) 
\end{equation}

where $\alpha^{intra}(i)$ and $\alpha^{inter}(i)$ refer to the $k$-shell value of node $i$ on the graphs constituting intra-community links and inter-community links, respectively. In this study, $\delta$ is set to 0.5 so that neither hubs nor bridges are preferentially selected over one another.

\section{Data and Tools}
\label{sec:DataToolsAndMethods}
This section briefly presents the networks used to conduct the study. It also reports on the SIR evaluation process and the measures used in the comparative analysis.

\subsection{Data}
The experiments concern fifteen real-world networks spanning biological networks, infrastructural networks, and offline/online social networks. Table \ref{TableBasicTopology}
reports their basic topological characteristics.

\textbf{Biological Networks}: In Yeast Protein  \cite{nr} and Yeast Collins \cite{netz}, the nodes represent proteins and are connected to each other if there's a physical interchange.

\textbf{Infrastructural Networks}: In the EU Airlines network \cite{netz} and U.S. Airports network \cite{kunegis2014handbook}, nodes represent European and U.S. airports, respectively. Nodes are connected if there's a direct flight between two airports. In the U.S. Power Grid network \cite{kunegis2014handbook}, nodes represent either a generator, transformer, or substation, and edges represent a power supply line.

\textbf{Offline Social Networks}: In the Adolescent Health network \cite{kunegis2014handbook}, nodes represent students and edges represent friendships. Students have been asked to list 5 of their female and male friends.

\textbf{Online Social Networks}: Six Facebook networks are under study (Facebook Friends, Ego Facebook, Caltech, Princeton, Facebook Organizations, and Facebook Politician Pages \cite{netz,nr}). In all networks, nodes represent Facebook users, and edges represent online friendships except for Facebook Politician Pages. In the latter, nodes represent politician pages from different countries, and edges represent mutual likes. In an online social pet platform called ``Hamsterster" \cite{kunegis2014handbook}, nodes represent users of this platform, and edges represent online friendships. The DeezerEU network \cite{rozemberczki2020characteristic} is formed based on the Deezer platform for music streaming. In this network, nodes are users from European countries, and edges represent mutual follower relationships. Finally, in DNC Emails \cite{netz}, nodes represent members of the Democrat National Committee, and edges represent email exchanges.

\begin{table}[t]
\caption{Symbols used in this study.}
\begin{center}
\begin{tabular}{|p{2cm} p{7cm}|}
\hline
Symbol & Meaning \\
\hline
$V$ & Set of nodes \\ 
$N$ & Total number of nodes \\
$E$ & Set of edges \\
$M$ & Total number of edges \\
$\varsigma$ & Set of communities \\
$C$ & Total number of communities \\
$c_q$ & $q$-th community \\
$n_{c_q}$ & Total number of nodes in community $c_q$ \\
$k_i^{intra}$ & Intra-community links of a node \\
$k_i^{inter}$ & Inter-community links of a node \\
$k_{i, c_q}$ & Total number of links of a node in community $c_q$ \\
$k_i^{tot}$ & Total number of links of a node \\
$a_{ij}$ & Connectivity between nodes $i$ and $j$ \\ 
$H_i$ & Entropy of the node's intra/inter-community  links \\
$\rho_i^{intra}$ & Fraction of intra-community links of a node \\
$\rho_i^{inter}$ & Fraction of inter-community links of a node \\
$\mu_{c_q}$ & Fraction of inter-community links in community $c_q$ \\
$\mu$ & Fraction of inter-community links in graph $G$\\
$R$ & A constant for standardization \\
$f_o$ & Fraction of initially infected nodes \\
$Q(G)$ & Newman's modularity \\
$Q(G \setminus \{i\})$ & Newman's modularity after the removal of node $i$ \\
$\alpha_{CHB}$ & Community Hub-Bridge \\
$\alpha_{PC}$ & Participation Coefficient \\
$\alpha_{CBM}$ & Community-based Mediator \\
$\alpha_{Comm}$ & Comm Centrality \\
$\alpha_{MV}^+$ & Modularity Vitality \\
$\alpha_{CBC}$ & Community-based Centrality \\
$\alpha_{ks}$ & K-shell with Community \\

\hline

\end{tabular}
\label{TableNotations}
\end{center}
\end{table}

\begin{table*}[t]
\caption{Main differences between the investigated community-aware centrality measures.}
\begin{center}
\begin{tabular}{|p{3.8cm} p{2cm} p{3.2cm} p{3.2cm}|}
\hline
Measure & Complexity & Description & Limitation(s) \\
\hline

Community Hub-Bridge  & $O(N^2C)$ & Detects hubs and bridges by giving preference to the local community size and direct belonging communities & May miss influential hubs if the network has a high number of communities \\ 

  &   &   &   \\ 

Participation Coefficient & $O(N<k>)$ & Finds nodes having a uniform link distribution across all the communities  & It indirectly prioritizes bridges over hubs and it may result in non-unique of centrality values \\

 &   &   &   \\ 

Community-based Mediator & $O(NM<k>)$ & Discovers intermediate nodes capable of quickly disseminating information across the communities & Reduces to degree centrality if the internal and external link density of a node are equal \\

&   &   &   \\ 

Comm Centrality & $O(NC)$ & Searches for hubs and bridges while taking into consideration the community structure strength & Directly prioritizes bridges and is undefined if at least one community is disconnected from the others\\

  &   &   &   \\ 

Modularity Vitality & $O(NC+M)$ & A signed measure that can be used to explicitly target hubs, bridges, or both & Susceptible to the limitations of Newman's modularity such as the resolution limit and local maxima \\

  &   &   &   \\

Community-based Centrality  & $O(N<k>)$ & Detects influential nodes based on the communities they participate in and their sizes  & The higher the number of communities, the more similar it is to normalized degree centrality \\

  &   &   &   \\ 

K-shell with Community & $O(NM)$& Determines core nodes embedded in local and global communities & Redundancy of nodes existing at the same hierarchical level\\

\hline

\end{tabular}
\label{TableCharacteristcs}
\end{center}
\end{table*}

\begin{table}[t]
\caption{Topological features of the networks. $N$ is the number of nodes. $M$ is the number of edges. $<k>$ is the average degree. $\varkappa$ is the transitivity. $Q$ is the modularity. $\lambda_{th}$ is the epidemic threshold. * means the largest connected component of the network is taken if it is disconnected.}
\begin{center}
\begin{tabular}{|p{4cm}|c|c|c|c|c|c|}
\hline
\textbf{Network} & $N$ & $M$ & $<k>$ & $\varkappa$ & $Q$ & $\lambda_{th}$ \\
\hline
A) Facebook Friends* & 329 & 1,954 & 11.88 & 0.51 & 0.69 & 0.05\\
B) EU Airlines & 417 & 2,953 &  14.16 & 0.30 & 0.11 & 0.02\\
C) U.S. Airports & 500 & 2,980 & 11.92 & 0.35 & 0.16 & 0.02 \\
D) Caltech* & 762 & 16,651 & 43.70 & 0.29 & 0.39 & 0.05 \\
E) DNC Emails* & 849 & 10,384 & 24.46 & 0.55 & 0.42 & 0.01 \\
F) Yeast Collins* & 1,004 &  8,319 & 16.57 & 0.62 & 0.75 & 0.03 \\
G) Yeast Protein* & 1,458 & 1,993 & 2.73 & 0.05 & 0.75 & 0.16 \\
H) Hamsterster* & 1,788  & 12,476 & 13.49 & 0.09 & 0.39 & 0.02\\
I) Adol. Health & 2,539 & 10,455 & 8.23 & 0.04 & 0.57 & 0.11 \\
J) Ego Facebook & 4,039 & 88,234 & 43.69 & 0.52 & 0.81 & 0.01 \\ 
K) U.S. Power Grid & 4,941 & 6,594 & 2.66 & 0.10 & 0.83 & 0.35 \\
L) Facebook Organizations & 5,524 & 94,219 & 34.11 & 0.22 & 0.59 & 0.02 \\
M) Facebook Politician Pages & 5,908 & 41,729 & 14.12 & 0.30 & 0.84 & 0.02 \\ 
N) Princeton* & 6,575 & 293,307 & 89.21 & 0.16 & 0.42 & 0.01\\
O) DeezerEU & 28,281  & 92,752 &  6.55 & 0.10 & 0.57 & 0.07 \\

\hline

\end{tabular}
\label{TableBasicTopology}
\end{center}
\end{table}

\subsection{Infomap Community Detection Algorithm}
We use Infomap \cite{rosvall2008maps} to uncover the communities of the real-world networks. Indeed, this popular algorithm has proved to be quite efficient in artificial and real-world benchmarks \cite{Yang2016, orman2011accuracy, orman2012comparative, jebabli2018community}. It minimizes the information flow used to characterize the connections in a network. Consider a random walker. It is more likely to stay longer in a densely connected cluster and rarely jumps to another cluster. Accordingly, Infomap uses Huffman coding to compress the encoded information of the random walker in each community. The optimal partitioning minimizes the map equation ($L$). It assigns nodes to clusters to reduce the random walker's movement. First, each node belongs to a unique community. Then, nodes are considered to be in the same community if they result in the largest decrease in $L$. This procedure iterates until the map equation reaches a minimum. In the end, one associates each node with a code in two parts: the first part (i.e., prefix) determines its community. The second part (i.e., codeword) refers to the node within the community.


\subsection{Kendall's Tau Correlation}
The Kendall's Tau correlation is used to measure the ordinal consistency of two sets of ranked nodes. Given two ranked sets $X=(x_{1}, x_{2}, ..,x_{n})$ and $Y=(y_{1}, y_{2}, ..., y_{n})$ of size $n$, a pair of ranks $(x_{i}, y_{i})$ and $(x_{j}, y_{j})$ is considered concordant if $x_{i} > x_{j}$ and $y_{i} > y_{j}$, or if $x_{i} < x_{j}$ and $y_{i} < y_{j}$. It is said discordant if $x_{i} > x_{j}$ and $y_{i} < y_{j}$, or if $x_{i} < x_{j}$ and $y_{i} > y_{j}$. In case of ties (if $x_{i} = x_{j}$ or $y_{i} = y_{j}$), the pair of ranks is neither concordant nor discordant. The Kendall's Tau correlation $\tau_b$ between two ranking sets $X$ and $Y$ of size $t$ is given by:

\begin{equation}
\tau_b(X,Y)=\frac{n_{co}-n_{di}}{\sqrt{(n_{co}+n_{di}+x)(n_{co}+n_{di}+y)}}
\label{eq2}
\end{equation}

where $n_{co}$ and $n_{di}$ denote the number of concordant and discordant pairs, respectively. $x$ and $y$ detain the number of tied pairs in sets $X$ and $Y$, respectively. $\tau_b$ provides a value in the interval [-1,1]. If $\tau_b > 0$, there is a positive monotonic association between the two sets. If $\tau_b < 0$, there is a negative monotonic association between the two sets.

\subsection{Pearson Correlation}
The Pearson correlation coefficient measures the linear association and direction between two variables. It is given by:

\begin{equation}
    \rho(X,Y) = \frac{\sum_{i=1}^{n}(X_i - \bar{X})(Y_i - \bar{Y} )}{\sqrt{\sum_{i=1}^{n}(X_i - \bar{X} )^{2} \sum_{i=1}^{n}(Y_i - \bar{Y})^{2}}}
\end{equation}
where $n$ is the sample size, $\bar{X}=\frac{\sum_i^n X_i}{n}$ is the sample mean of variable $X$, and $\bar{Y}=\frac{\sum_i^n Y_i}{n}$ is the sample mean of variable $Y$. Pearson's correlation values are in the range [-1,+1]. The greater the absolute value of the correlation coefficient, the stronger the relationship. Extreme values indicate a perfect linear relationship. The sign of the correlation coefficient gives the direction of the relationship. It is positive when the value of one variable increases, the value of the other variable also tends to increase. Inversely, negative coefficients represent cases when the value of one variable increases and the other variable's value tends to decrease.

\subsection{Susceptible-Infected-Recovered Model}
\label{sec:SIR}
The Susceptible-Infected-Recovered (SIR) model \cite{anderson1979population} is a popular diffusion model for assessing the effectiveness of centrality measures. It can be used to model a wide range of real-world scenarios such as viral marketing and cyberattacks and has numerous implications \cite{ismail2017susceptible, nguyen2017modelling, buckee2021thinking, toda2020susceptible}. At first, one ranks the nodes in the decreasing order of their centrality value. A given proportion ($f_o$) of the top-ranked nodes is set in the infectious state (I). All the other nodes are in the susceptible state (S). An infectious node infects its susceptible neighbors with a probability $\lambda$, as the propagation proceeds. At the same time, an infected node can recover at a rate $\gamma$. By the end of the propagation phenomenon, the disease propagation stops when all nodes are either in the susceptible or recovered state. Here, the outbreak size based on the number of nodes in the recovered state (R) is measured. The outbreak size determines the spreading power of $f_o$. Therefore, the goal is to maximize this value. All networks have an epidemic threshold that controls the epidemic spreading. In this study, we use the following definition to calculate it \cite{wang2016predicting}:
\begin{equation}
\lambda_{th} = \frac{<k>}{<k^2> -  <k>}
\end{equation}
where $<k>$ and $<k^2>$ are the first and second moments of the network's degree distribution. The epidemic threshold value of each network is provided in Table \ref{TableBasicTopology}. Note that we use three values of the infection rate in the experiments (the epidemic threshold $\lambda_{th}$,  $\lambda_{th}/2$ and $\lambda_{th} \times 2$). As we do not observe significant differences, we report only the results equal to the epidemic threshold. Moreover, the SIR simulations are averaged over 100 independent iterations for each network.


\subsection{The Mixing Parameter} 
The mixing parameter characterizes how strong or weak a community structure is. If the network has well-separated communities, its mixing parameter is low, indicating a strong community structure. If the network's communities are not clearly defined, the mixing parameter is high, indicating a weak community structure. It is quantified by the fraction of inter-community links over the total links in a network. Its value is in the range of [0,1]. The less the inter-community links (i.e., links from one community to another, the stronger the community structure is. It is given by:

\begin{equation}
\mu =
\frac{\sum_{i=1}^{N} k_i^{inter}}{\sum_{i=1}^{N} k^{tot}_i}
\end{equation}

where $\sum_{i=1}^{N}k_i^{inter}$ represents the inter-community links of all the nodes in the network and $\sum_{i=1}^{N} k^{tot}_i$ represents the total sum of all degrees in the network.

\subsection{Evaluation Criterion}
We use the degree centrality as a baseline to compare the spreading outbreak size of community-aware centrality measures. Indeed, all the community-aware centrality measures are extensions of degree centrality ($k_i^{tot}$ = $k_i^{intra}$ + $k_i^{inter}$) which in turn is agnostic to community structure. The relative difference is defined as:
\begin{equation}
\Delta R = \frac{R_c - R_b}{R_b}
\end{equation}


where $R_c$ denotes the total number of recovered nodes using a specific community-aware centrality measure $c$. $R_b$ represents the total number of recovered nodes with the baseline degree centrality. A positive value of $\Delta R$ indicates that the community-aware centrality measure is more effective than the baseline.

\section{Experimental Results}
This section reports the results of the experiments. First of all, we examine how the various community-aware centrality measures correlate. Then, we investigate the influence of the network's community structure strength on the correlation. Finally, we compare the spreading effectiveness of the centrality measures.

\label{sec:ExpResults}
\subsection{Correlation of the Community-aware Centrality Measures}
\begin{figure}[ht!]
\centerline{\includegraphics[width=3.6 in, height=3.2 in]{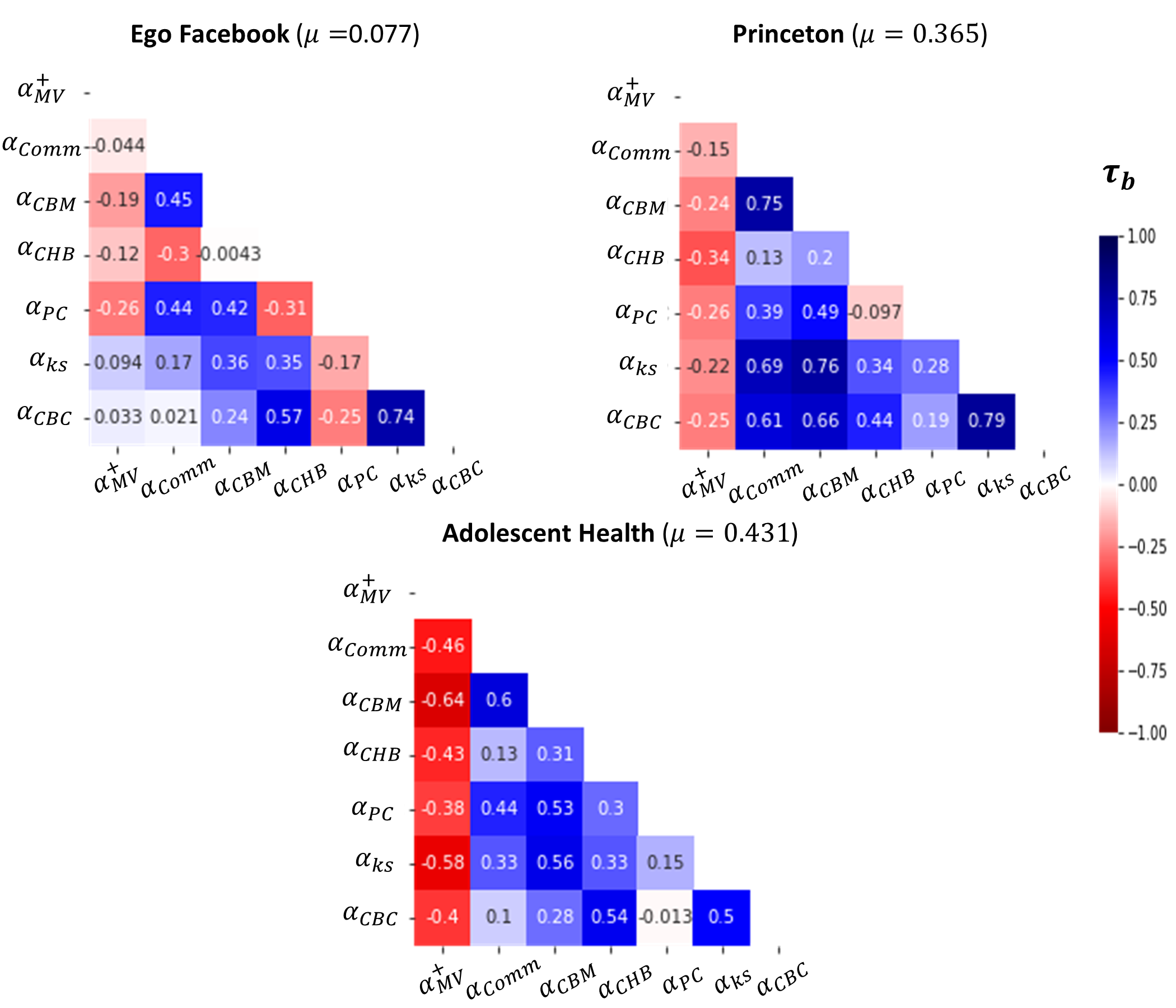}}
\caption{Kendall's Tau correlation ($\tau_b$) between all the community-aware centrality measures:   Modularity Vitality = $\alpha^+_{MV}$, Comm Centrality = $\alpha_{Comm}$, Community‑based Mediator = $\alpha_{CBM}$, Community Hub-Bridge = $\alpha_{CHB}$, Participation Coefficient = $\alpha_{PC}$, K-shell with Community = $\alpha_{ks}$, and Community-based Centrality = $\alpha_{CBC}$.}
\label{Fig1A-CorrelationOf3NetworksSeparately}
\end{figure}

The first experiment investigates the correlation between the various couples of community-aware centrality measures in each network. Figure \ref{Fig1A-CorrelationOf3NetworksSeparately} illustrates the correlation heatmaps for three typical networks. The heatmaps of the other networks are available in the supplementary material.

The heatmap of the first typical network, Ego Facebook, is very patchy. In other words, the correlation between the various community-aware centrality measure ranges from low negative correlation to high positive values. For example, Community-based Centrality has a negative correlation value ($-$0.25) with Participation Coefficient and a high positive value ($+$0.74) with K-shell with Community. Similar behavior characterizes EU Airlines, U.S. Airports, Facebook Politician Pages, Facebook Friends, Yeast Collins, and U.S. Power Grid.

The second typical network is Princeton. Modularity Vitality has a weak negative or positive correlation with its alternatives. The correlation values are more uniform between the other community-aware centrality measure. Note, however, that some low negative correlation values may appear. Networks that follow this behavior are DNC Emails, Yeast Protein, Hamsterster, and Facebook Organizations.

The third typical network is Adolescent Health. Correlation of Modularity Vitality exhibits weak to medium negative correlation values with the others. The remaining correlation values are positive, ranging from low to medium values. One observes similar behavior in Caltech and DeezerEU.

\begin{figure*}[ht!]
\centerline{\includegraphics[width=6.2 in, height=2.2 in]{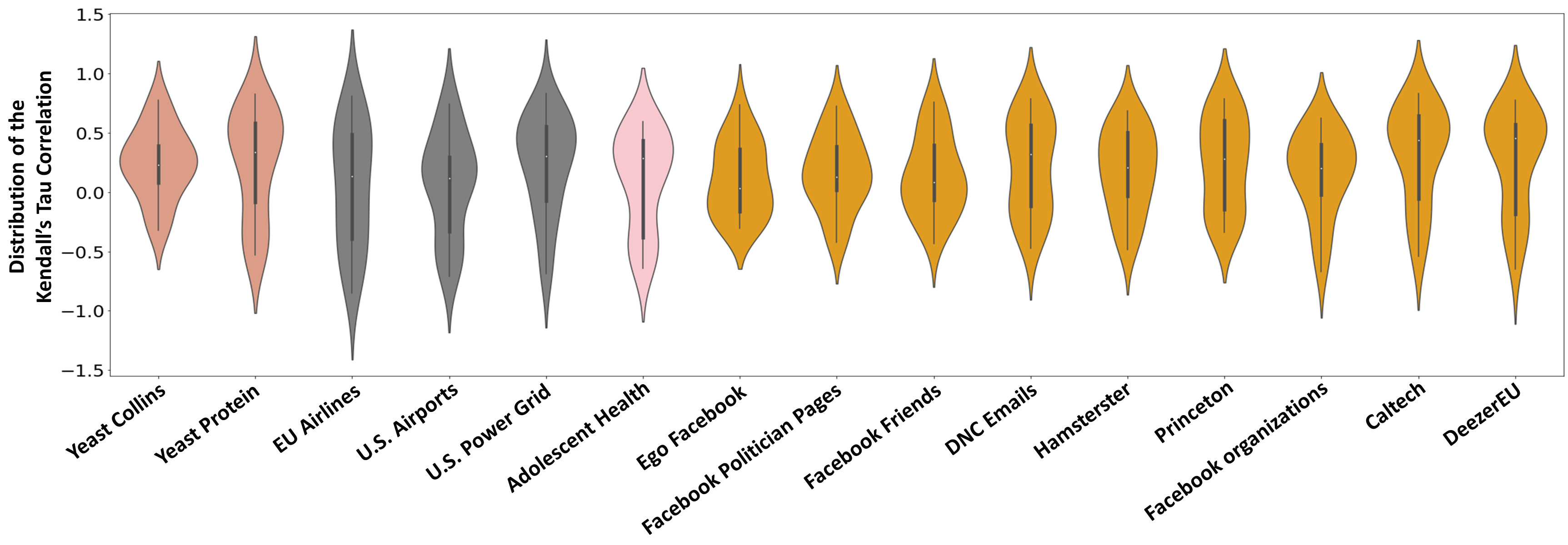}}
\caption{Violin plots of Kendall's Tau correlation between the community-aware centrality measures for each network. Colors represent the network's domain. Biological networks are dark pink.  Infrastructural networks are grey. Offline social networks are light pink. Online social networks are orange.}
\label{FigViolin}
\end{figure*}
Figure \ref{FigViolin} shows the distribution of Kendall's Tau correlation between the community-aware centrality measures for each network. Note that some distributions may pass the extremes [-1, +1] due to the smoothing of the kernel density function. One observes significant variation of the distributions' range for biological networks and online/offline networks. For example, the distribution of Yeast Collins is more compact as compared to Yeast Protein. Similar observations apply to Ego Facebook compared to DeezerEU. Additionally, three networks have bimodal distributions (EU Airlines, DNC Emails, and Princeton). The first modal value is positive, and the second is negative. The remaining network distributions are unimodal. That being said, the average mean of networks' correlation distributions is 0.19$\pm$0.08 and the average median is 0.24$\pm$0.12. Hence, the correlation between the community-aware centrality measures is globally low. Additionally, there is an emerging pattern related to Modularity Vitality. A closer inspection of its correlation values shows that the negative tail of the distribution is generally related to the combinations including Modularity Vitality.

Hence, Modularity Vitality is extracting diverging information with the alternative community-aware centrality measures. The negative correlation values are not unexpected. Indeed, Modularity Vitality favors hubs while the other community-aware centrality measures mainly pinpoint bridges as the most influential nodes. Nonetheless, even if they aim at targeting bridges, it does not mean that they are highly correlated.

\begin{figure}[ht!]
\centerline{\includegraphics[width=3.8 in, height=3.2 in]{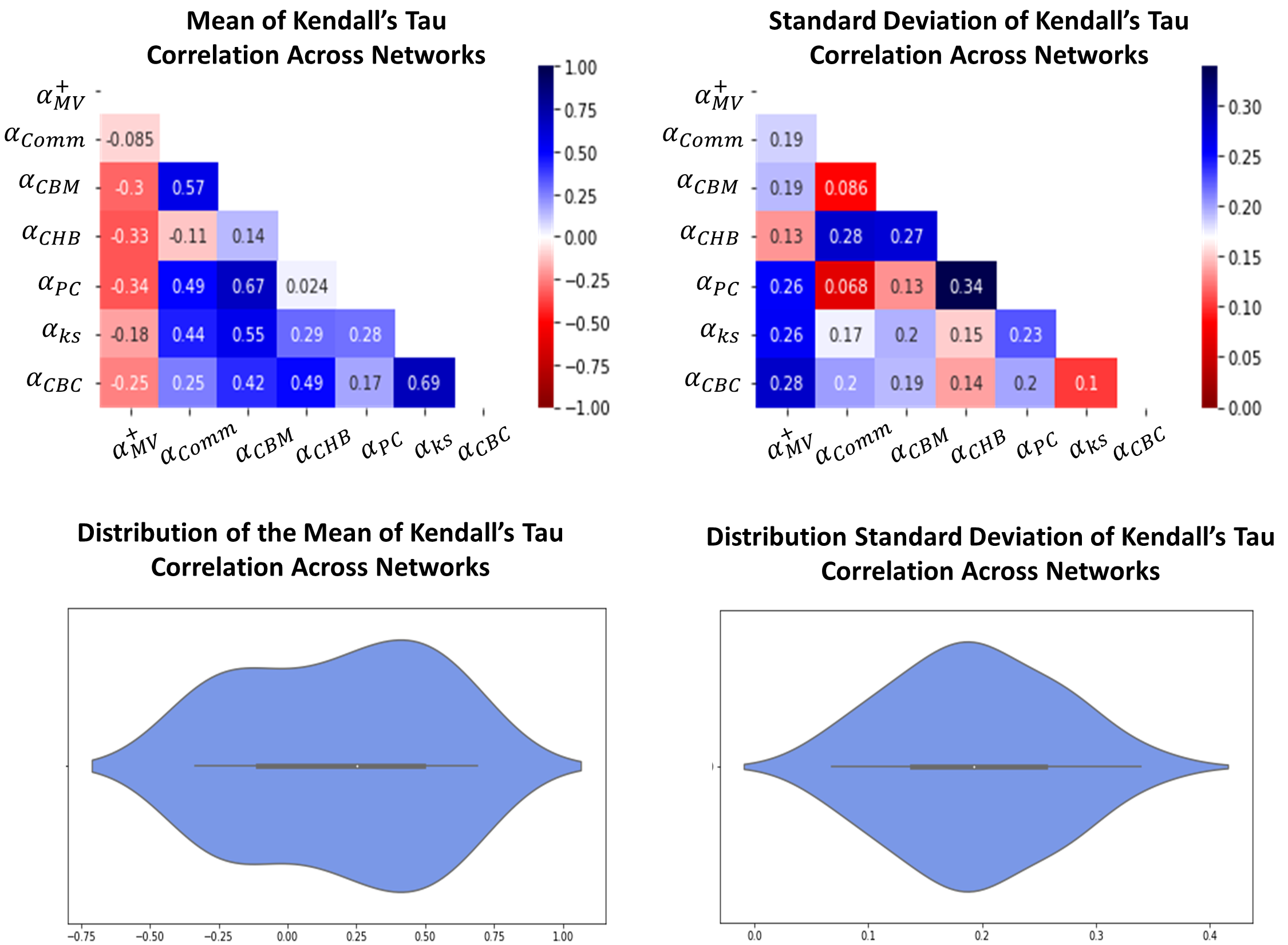}}
\caption{Upper figures: Mean and standard deviation of Kendall's Tau correlation between the community-aware centrality measures across the fifteen real-world networks. Bottom figures: The distribution of the mean and standard deviation of Kendall's Tau correlation between the community-aware centrality measures across the fifteen real-world networks. The community-aware centrality measures are:  Modularity Vitality targeting hubs = $\alpha^+_{MV}$, Comm Centrality = $\alpha_{Comm}$, Community‑based Mediator = $\alpha_{CBM}$, Community Hub-Bridge = $\alpha_{CHB}$, Participation Coefficient = $\alpha_{PC}$, K-shell with Community = $\alpha_{ks}$,  and Community-based Centrality = $\alpha_{CBC}$.}
\label{Fig1C-CorrelationAcrossNetworks}
\end{figure}

We calculate the mean and the standard deviation of the Kendall's Tau correlation between the community-aware centrality measures across the fifteen networks. This enables us to investigate if they generally tend to correlate or not independently of the network structure. The top left figure in Fig. \ref{Fig1C-CorrelationAcrossNetworks} shows two emerging patterns. The first concerns Modularity Vitality. It has a weak negative correlation with the rest of the community-aware centrality measures. The couple consisting of Community Hub-Bridge with Comm Centrality also shows a weak negative correlation. The second concerns the other community-aware centrality measures. They present low to medium positive correlation values. Indeed, whatever the combination, correlation is always smaller than 0.70.

The bottom left figure shows the distribution of the top left figure in in Fig. \ref{Fig1C-CorrelationAcrossNetworks}. It is bimodal, with a positive mode and a negative one. This observation corroborates previous results on the distribution of each network reported in Fig. \ref{FigViolin}. It further explains the two emergent patterns encountered with the community-aware centrality measures under test.

The top right figure in Fig. \ref{Fig1C-CorrelationAcrossNetworks} shows the standard deviation of the means across networks. It ranges from 0.068 to 0.34. Its most frequent value is around 0.2. The violin plot reported in the bottom right figure presents its distribution. The mean value is 0.19$\pm$0.07. Here, the distribution is symmetric and unimodal.

To summarize, these experiments show that one can consider three typical behaviors for the correlation heatmaps of the community-aware centrality measures. These cases span from low correlation to high correlation. However, generally, the correlation between community-aware centrality measures ranges from low to medium. The values can mainly be negative when Modularity Vitality is involved or positive for the other community-aware centrality measures, which target mainly bridges.

\subsection{Influence of the Community Structure Strength on Correlation}
The previous experiment shows that one can distinguish three groups of networks according to the visual similarity of their heatmaps. To further this analysis, we associate each network to a vector made of its set of Kendall's Tau correlation values. Afterward, we compute the Pearson correlation between the couples of vectors.

\begin{figure}[h]
\centerline{\includegraphics[width=0.8\linewidth, height=5 in]{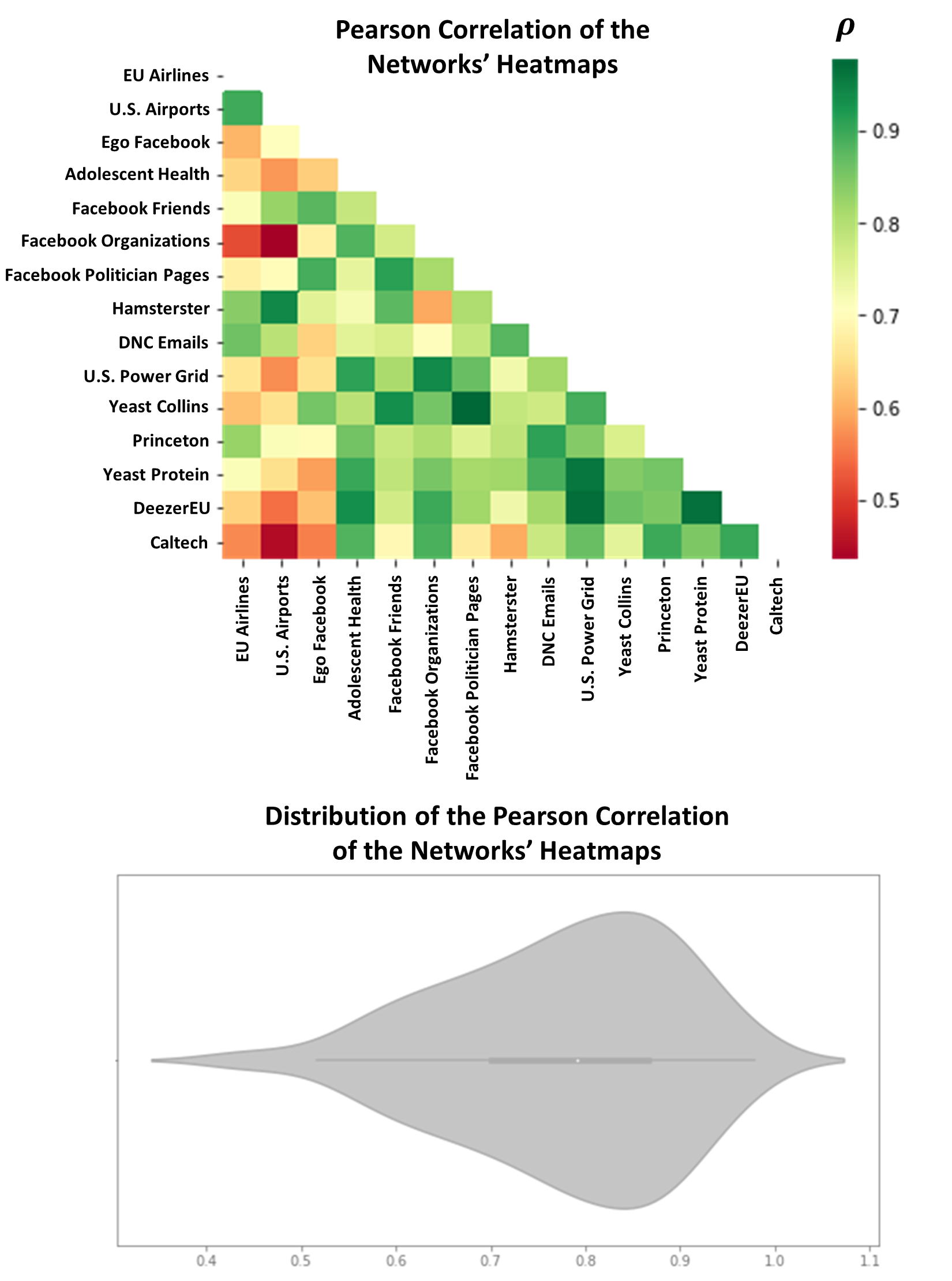}}
\caption{Top figure: Pearson Correlation ($\rho$) between networks based on their Kendall's Tau correlation values where each network is associated to its Kendall's Tau correlation vector. Bottom figure: Violin plot of the Pearson correlation.}
\label{Fig1B-CorrelationOfNetworks}
\end{figure}

Figure \ref{Fig1B-CorrelationOfNetworks} reports the heatmap of the Pearson correlation between the fifteen networks and its violin plot. Networks are ordered based on their mean correlation with other networks in increasing order. One can see that EU Airlines, U.S. Airports, and Ego Facebook are the least correlated with other networks, while Caltech, DeezerEU, and Yeast Protein are the most correlated. Inspecting the violin plot shows that predominantly the networks correlate well (with a mean of 0.77$\pm$0.12). Nevertheless, the left tail of the distribution indicates the presence of low correlation values.

A closer inspection of the network topological properties shows that networks that do not correlate well with the others have a strong community structure. In contrast, the most correlated to their pairs have a medium to weak community structure. It suggests that the community structure strength is playing a significant role in this behavior.

\begin{figure}[t]
\centerline{\includegraphics[width=2 in, height=1.5 in]{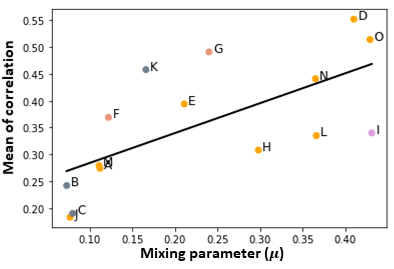}}
\caption{Linear regression between the mean of correlation of the community-aware centrality measures for each network and their mixing parameter ($\mu$). Ordinary least squares estimates are used.}
\label{Fig3-LinearReg}
\end{figure}

We perform a simple linear regression to relate the community structure strength with the mean correlation value associated with each network. Note that we discard Modularity Vitality in the mean calculation. Indeed, it exhibits a negative correlation with the other community-aware centrality measures. Hence, it may cancel out the contributions of the other community-aware centrality measures to the final mean correlation. The linear regression process uses ordinary least squares estimators. The mean values are the dependent variables, and the mixing parameter quantifying the community structure strength is the independent variable. The relationship between the dependent and independent variables is statistically significant when the $p$-value is below a threshold value.  
Figure \ref{Fig3-LinearReg} reports the results of the linear regression.  The linear relation hypothesis is statistically significant ($p < $ 0.01). The predicted slope is positive ($+$0.557). In other words, as the mixing parameter increases (i.e., as the network community structure gets weaker), the mean correlation tends to increase. 

We also perform a linear regression analysis to study the relationship between the number of communities in each network and the correlation of community-aware centrality measures. As results show no significant relationship, they are not reported here.

\begin{figure*}[ht!]
\centerline{\includegraphics[width=7 in, height=6 in]{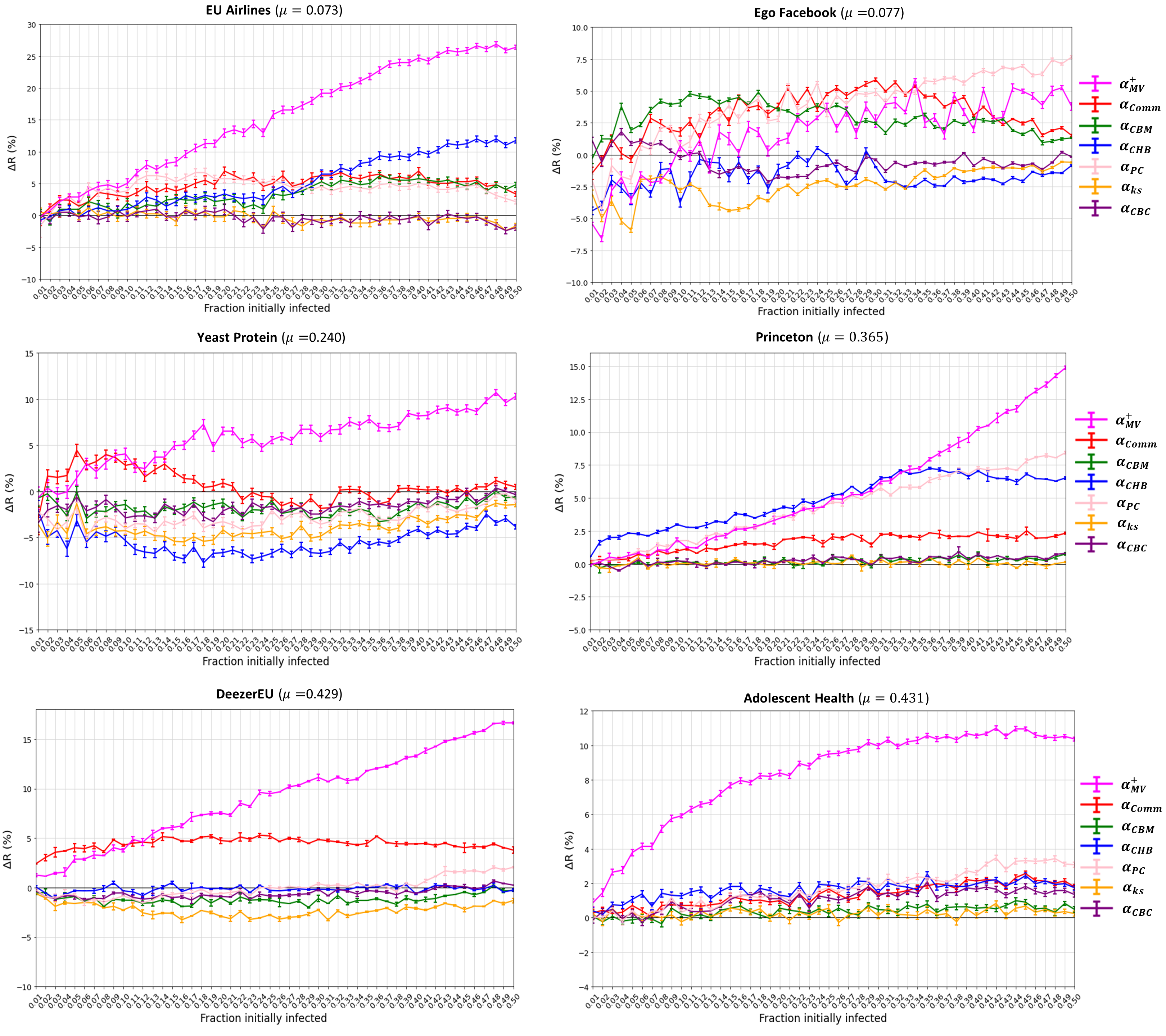}}
\caption{Relative difference of the outbreak size ($\Delta R$) as a function of the fraction of initially infected nodes for three real-world networks. The initial spreaders set is build according to the ranks associated to a given community-aware centrality measure. The community-aware centrality measures are:  Modularity Vitality targeting hubs = $\alpha^+_{MV}$, Comm Centrality = $\alpha_{Comm}$, Community‑based Mediator = $\alpha_{CBM}$, Community Hub-Bridge = $\alpha_{CHB}$, Participation Coefficient = $\alpha_{PC}$, K-shell with Community = $\alpha_{ks}$,  and Community-based Centrality = $\alpha_{CBC}$.}
\label{Fig2-SIR}
\end{figure*}

\subsection{Diffusion Effectiveness of the Community-aware Centrality Measures}

The diffusion performance of the community-aware centrality measures is studied using the SIR simulation process. One can consider any centrality measure as a baseline. In this study, we decide to use the classical degree centrality. Indeed, all the community measures are correlated with degree centrality, which is the sum of the total intra-community and inter-community links of the node ($k_i^{tot}$ = $k_i^{intra}$ + $k_i^{inter}$). Nonetheless, they are more complex as they exploit the mesoscopic information distinguishing hubs and bridges. Thus, a negative value indicates that the classical degree centrality is more efficient than the community-aware centrality measure under test to diffuse information in the network. In contrast, a positive value illustrates the benefits of community-aware centrality measures over the classical degree centrality measure.

To present the results of these experiments, we choose to consider three categories of networks according to the community structure strength. The first category regroups the networks with a strong community structure ($\mu \leq$ 0.084). It contains Ego Facebook, EU Airlines, and U.S. Airports. Networks with a medium community structure strength (0.084 $<\mu\leq$ 0.366) make the second category. It is the biggest one with Facebook Politician Pages, Facebook Friends, Yeast Collins, U.S. Power Grid, Princeton, DNC Emails, Yeast Protein, Hamsterster, and Facebook Organizations. Finally, the networks with a weak community structure ($\mu \geq$ 0.410) form the last category. It contains Caltech, Deezer EU, and Adolescent Health. Figure \ref{Fig2-SIR} shows the relative difference of the outbreak size ($\Delta R$) as a function of the fraction of initially infected nodes for six networks belonging to the three categories. Results for the other networks are available in the supplementary material.

\subsubsection{Networks with a Strong Community Structure}
Two typical behaviors emerge in networks with a strong community structure. The first is illustrated by EU Airlines and Ego Facebook represents the second. In EU Airlines, Modularity Vitality outperforms the rest of the centrality measures. Its gain over the baseline reaches 27\%. It is 15\% higher than Community Hub-Bridge, which ranks second. The U.S. Airports network shows similar behavior. Indeed, Modularity Vitality has the highest epidemic outbreak. Ego Facebook is a typical illustration of the second behavior. There is no single winner. The relative effectiveness of the community-aware centrality measures is tied to the proportion of initially infected nodes. In Ego Facebook, Community-based Mediator exhibits the highest outbreak difference for a low range of initially infected nodes. Then Comm Centrality takes the lead in the medium range. Finally, Participation Coefficient performs better in the last interval.

\subsubsection{Networks with a Medium Community Structure}
 

One can distinguish two cases in networks with a medium community structure strength. Yeast Protein network illustrates the first. It shows that at a small fraction of initially infected nodes, Comm Centrality outperforms the other community-aware centrality measures. Then Modularity Vitality takes over. Facebook Politician Pages, Facebook Friends, U.S. Power Grid, and Yeast Collins follow this behavior. The gain of Modularity Vitality over the baseline can reach up to 19\% (in Facebook Friends). Note that Comm Centrality stays the most performing at all fraction of initially infected nodes for the Yeast Collins network. Princeton represents the second case. It shows that at first, Community Hub-Bridge has the highest epidemic outbreak, and then Modularity Vitality which targets hubs, takes over. One observes a similar behavior in DNC Emails and Hamsterster. It is worth noticing that Modularity Vitality is the most effective at all fraction of initially infected nodes in Facebook Organizations network.

\subsubsection{Networks with a Weak Community Structure}

One can also notice two typical behaviors in networks with a weak community structure strength. The first one illustrated with DeezerEU shows that at a small fraction of initially infected nodes, Comm Centrality outperforms its alternatives. Then, Modularity Vitality takes over. Its gain reaches 17\%. It is 13\% higher than Comm Centrality that ranks second. Adolescent Health illustrates the second case. Modularity Vitality outperforms the other community-aware centrality measures at all fractions of initially infected nodes. Caltech also shows similar behavior.

\section{Discussion}
\label{sec:Discussion}

This study conducts a systematic evaluation of prominent community-aware centrality measures using a set of fifteen real-world networks spanning a wide variety of complex network structural patterns. First, we compute Kendall's Tau correlation between the community-aware centrality measures on each network. Results indicate that the correlation spans from low to medium. Moreover, one can arrange the community-aware centrality measures in two groups. The first group contains Modularity Vitality. A negative correlation with all the other community-aware centrality measures characterizes it. The remaining centrality measures form the second group. Correlation between these centrality measures ranges from low to medium positive values.  These two groups seem to reflect the ability of community-aware centrality measures to prioritize either hubs or bridges.  It all depends on how they integrate the intra-community and inter-community links.

\begin{figure*}[h!]
\centering
\begin{subfigure}{0.49\linewidth}
\centering
\includegraphics[height=6cm, width=6cm]{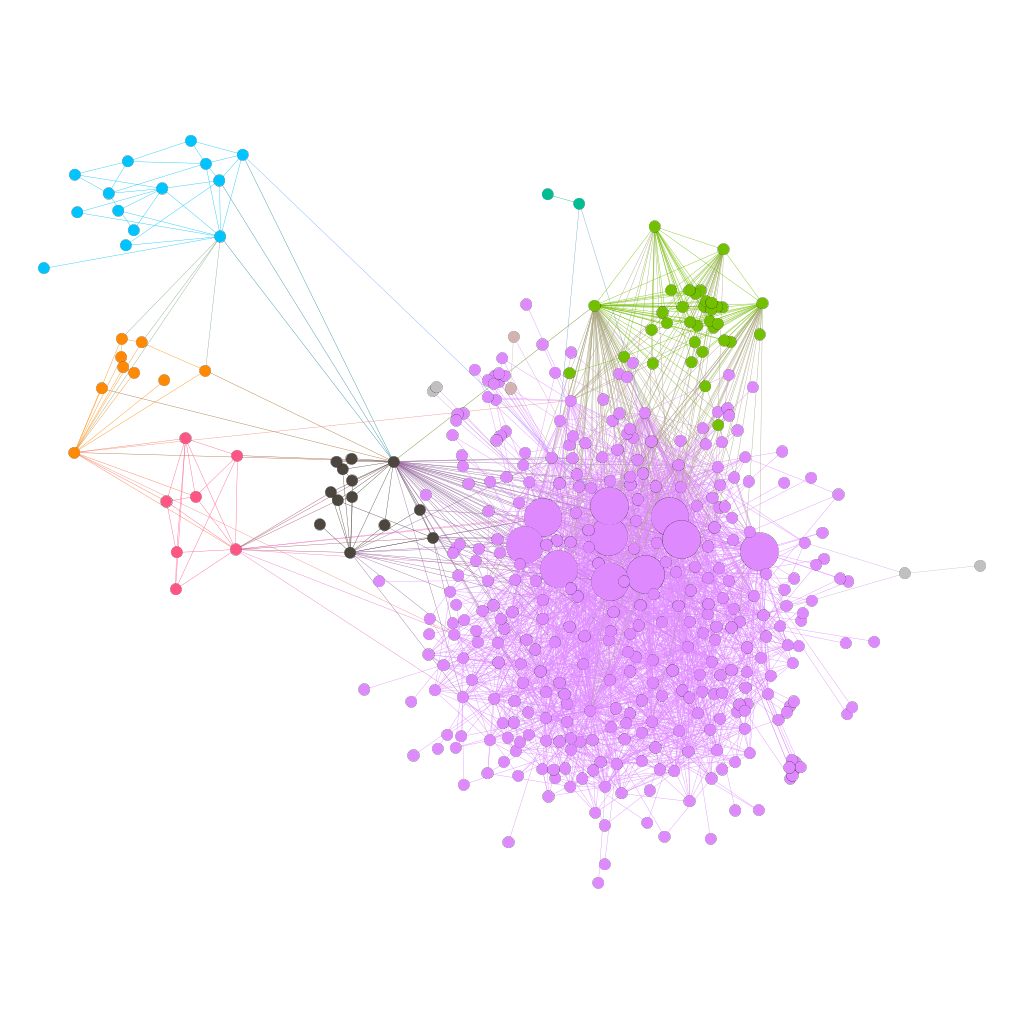}
    \caption{Nodes infected by K-shell With Community at $f_o$ = 1\%}
    \label{fig:my_sub_1_EUAirlines}
\end{subfigure}
\hfill
\begin{subfigure}{0.49\linewidth}
\centering
\includegraphics[height=6cm, width=6cm]{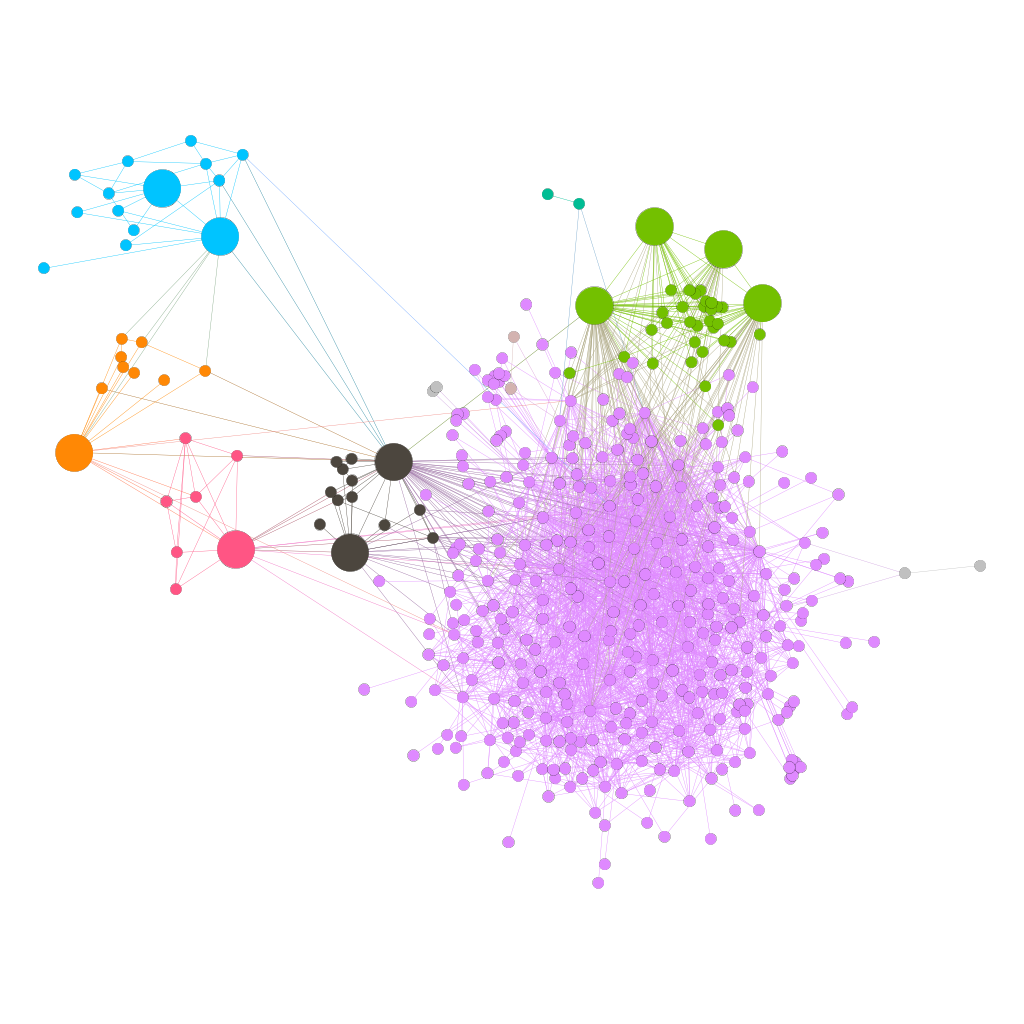}
    \caption{Nodes infected by Modularity Vitality at $f_o$ = 1\%}
    \label{fig:my_sub_2_EUAirlines}
\end{subfigure}
\vspace{0.5in} 
\begin{subfigure}{0.49\linewidth}
\centering
\includegraphics[height=6cm, width=6cm]{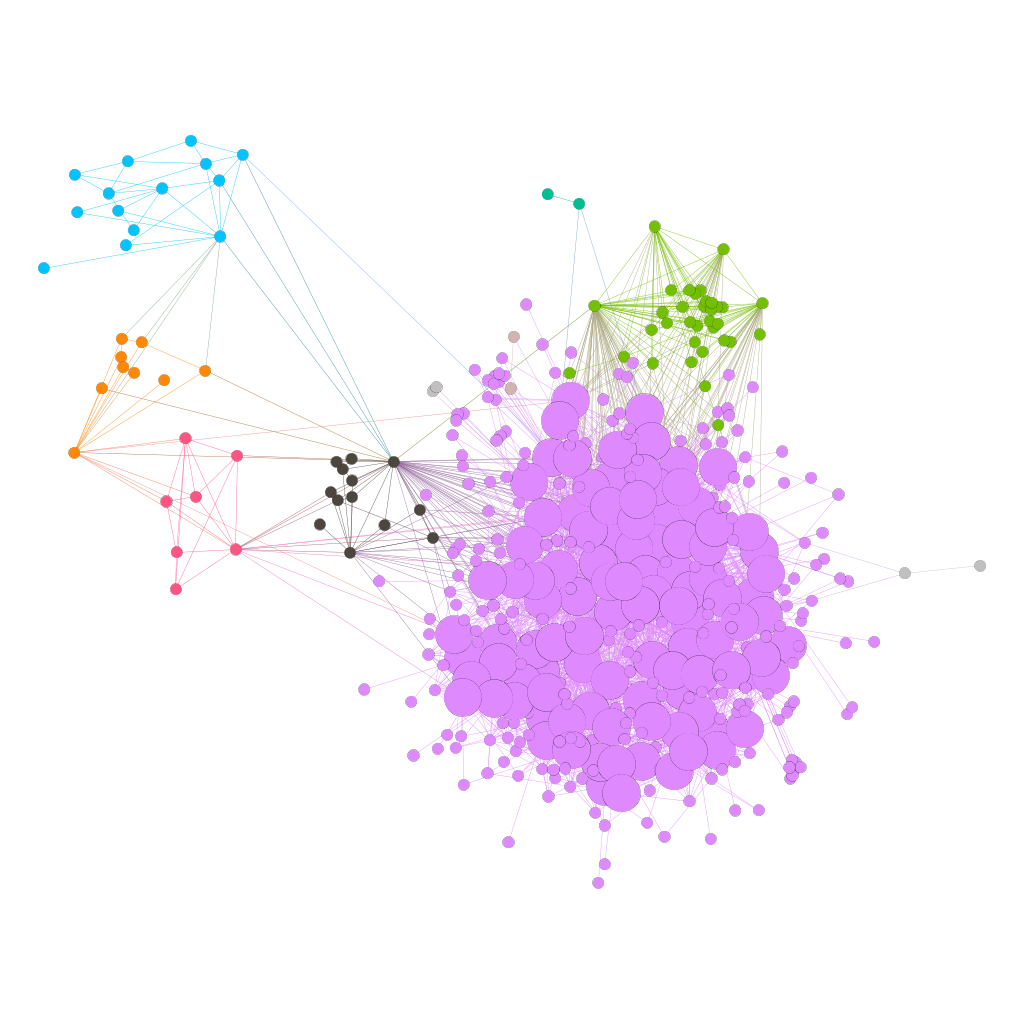}
    \caption{
    Nodes infected by K-shell with Community at $f_o$ = 25\%}
    \label{fig:my_sub_3_EUAirlines}
\end{subfigure}
\hfill
\begin{subfigure}{0.49\linewidth}
\centering
\includegraphics[height=6cm, width=6cm]{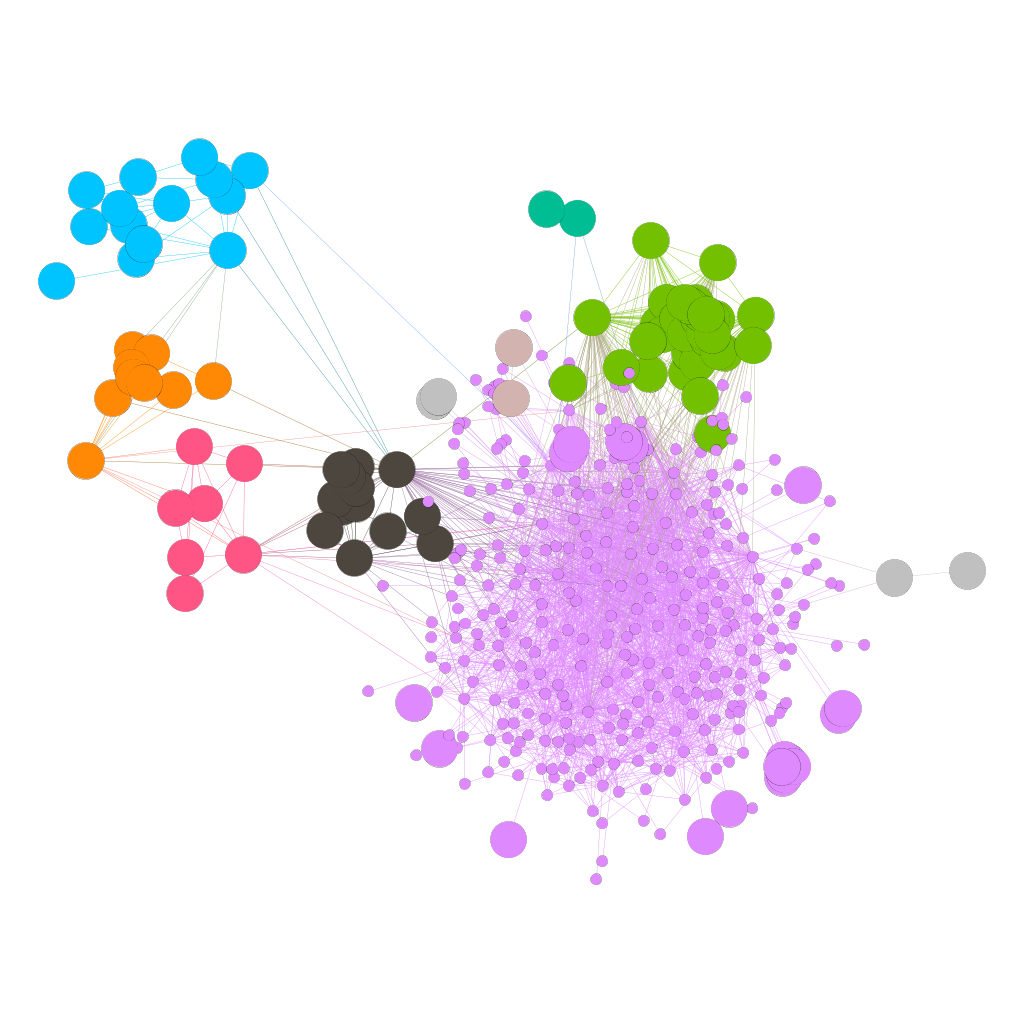}
    \caption{Nodes infected by Modularity Vitality at $f_o$ = 25\%}
    \label{fig:my_sub_4_EUAirlines}
\end{subfigure}
\caption{Comparing the position of the nodes infected at low availability of resources, at $f_o$ = 1\% (top figures) and at high availability of resources, at $f_o$ = 25\% (bottom figures) in the EU Airlines network. The bigger nodes in the left figure are the nodes picked by K-shell with Community and the bigger nodes in right figure are the nodes picked by Modularity Vitality.}
\label{EUAirlinesVisualization}
\end{figure*}

\begin{figure*}[h!]
\centering
\begin{subfigure}{0.49\linewidth}
\centering
\includegraphics[height=6cm, width=6cm]{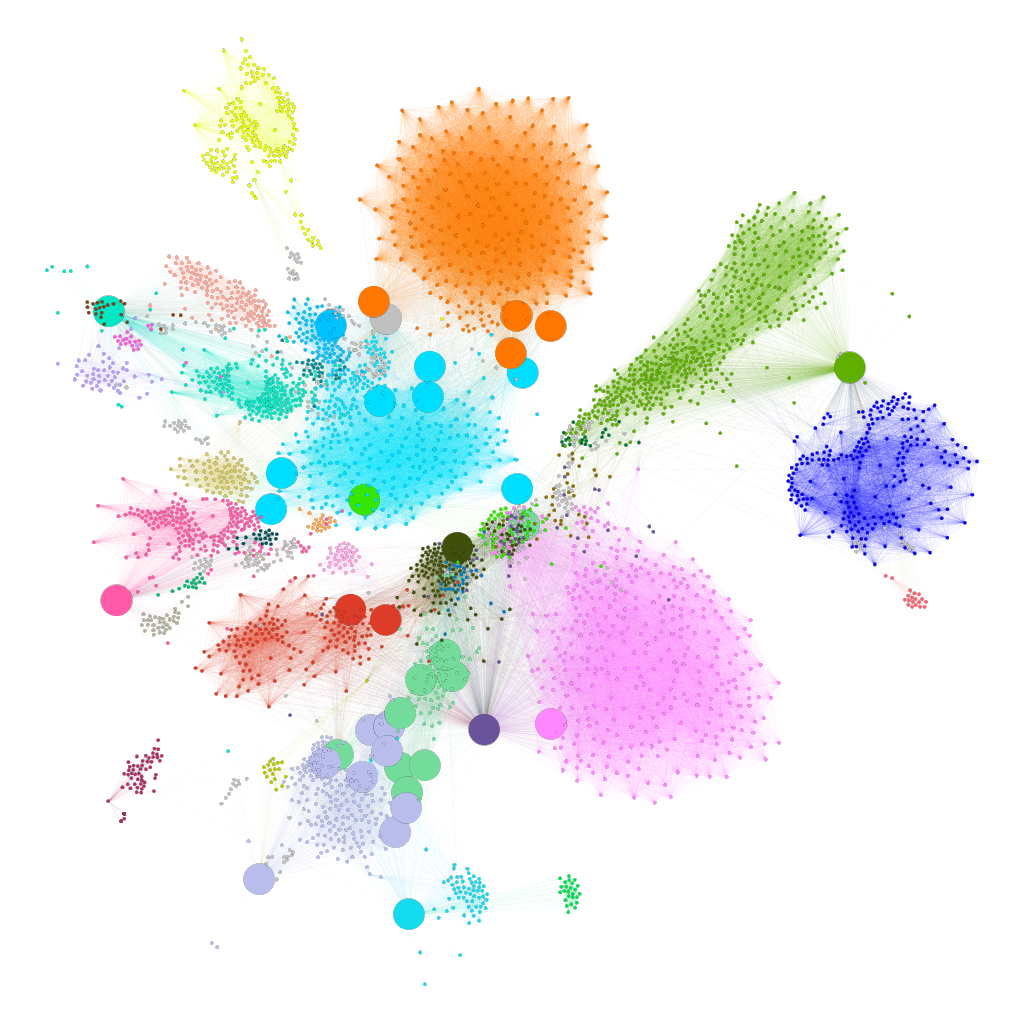}
    \caption{Nodes infected by Community-based Mediator at $f_o$ = 1\%}
    \label{fig:my_sub_1_EgoFacebook}
\end{subfigure}
\hfill
\begin{subfigure}{0.49\linewidth}
\centering
\includegraphics[height=6cm, width=6cm]{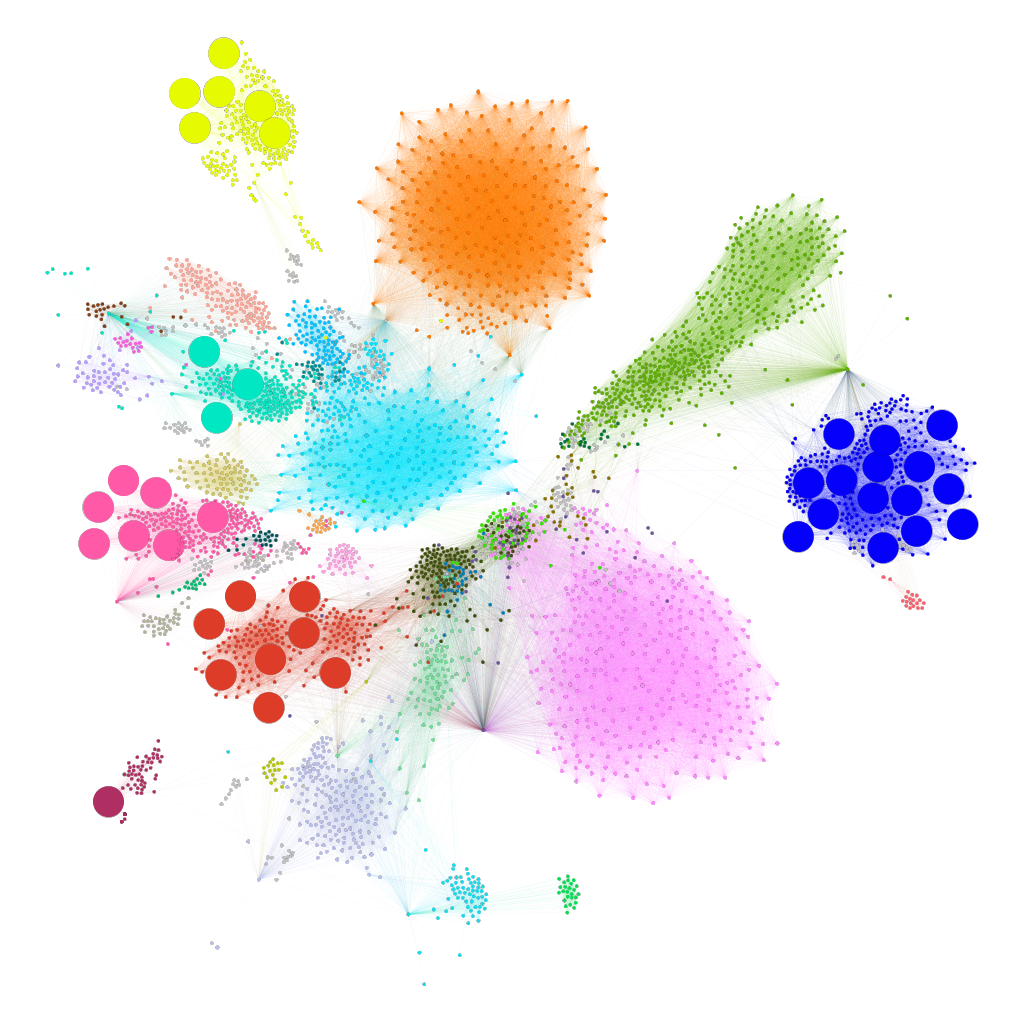}
    \caption{Nodes infected by Modularity Vitality at $f_o$ = 1\%}
    \label{fig:my_sub_2_EgoFacebook}
\end{subfigure}
\vspace{0.5in} 
\begin{subfigure}{0.49\linewidth}
\centering
\includegraphics[height=6cm, width=6cm]{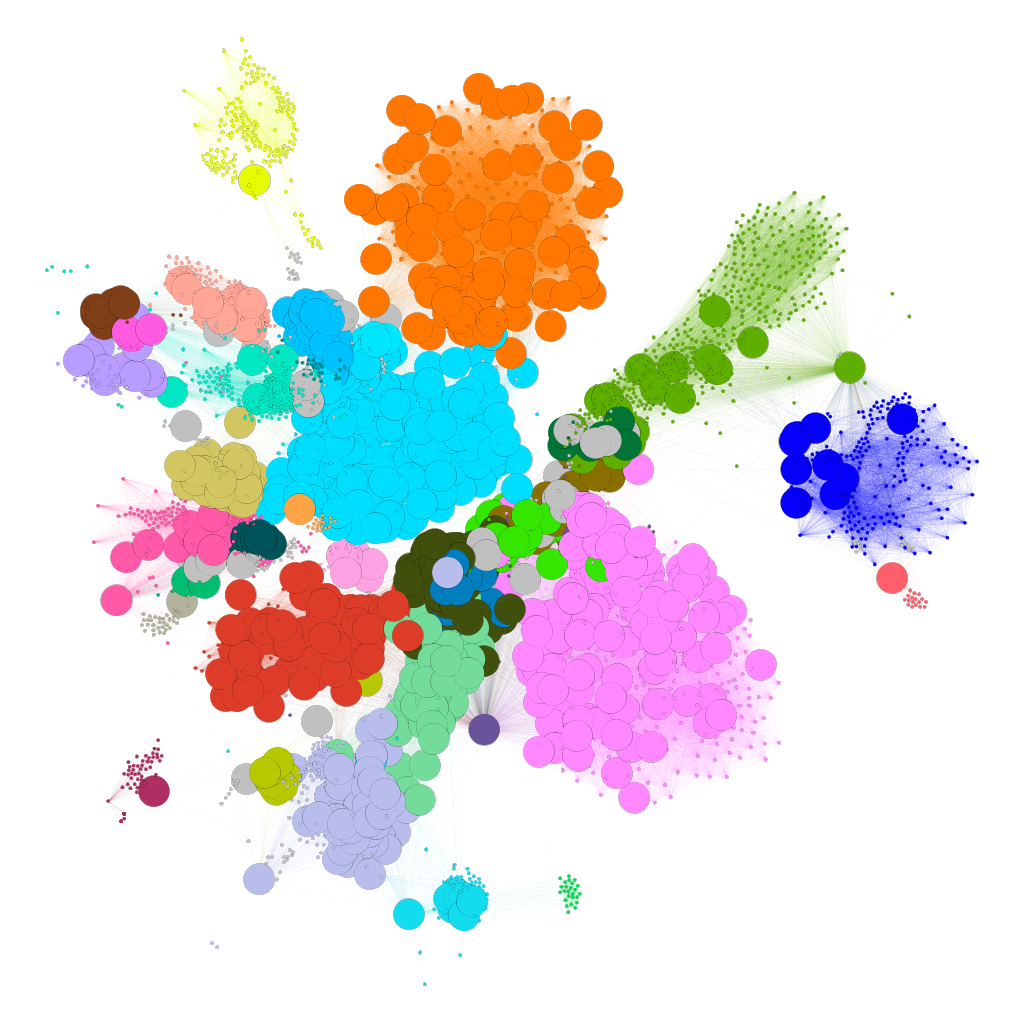}
    \caption{
    Nodes infected by Community-based Mediator at $f_o$ = 25\%}
    \label{fig:my_sub_3_EgoFacebook}
\end{subfigure}
\hfill
\begin{subfigure}{0.49\linewidth}
\centering
\includegraphics[height=6cm, width=6cm]{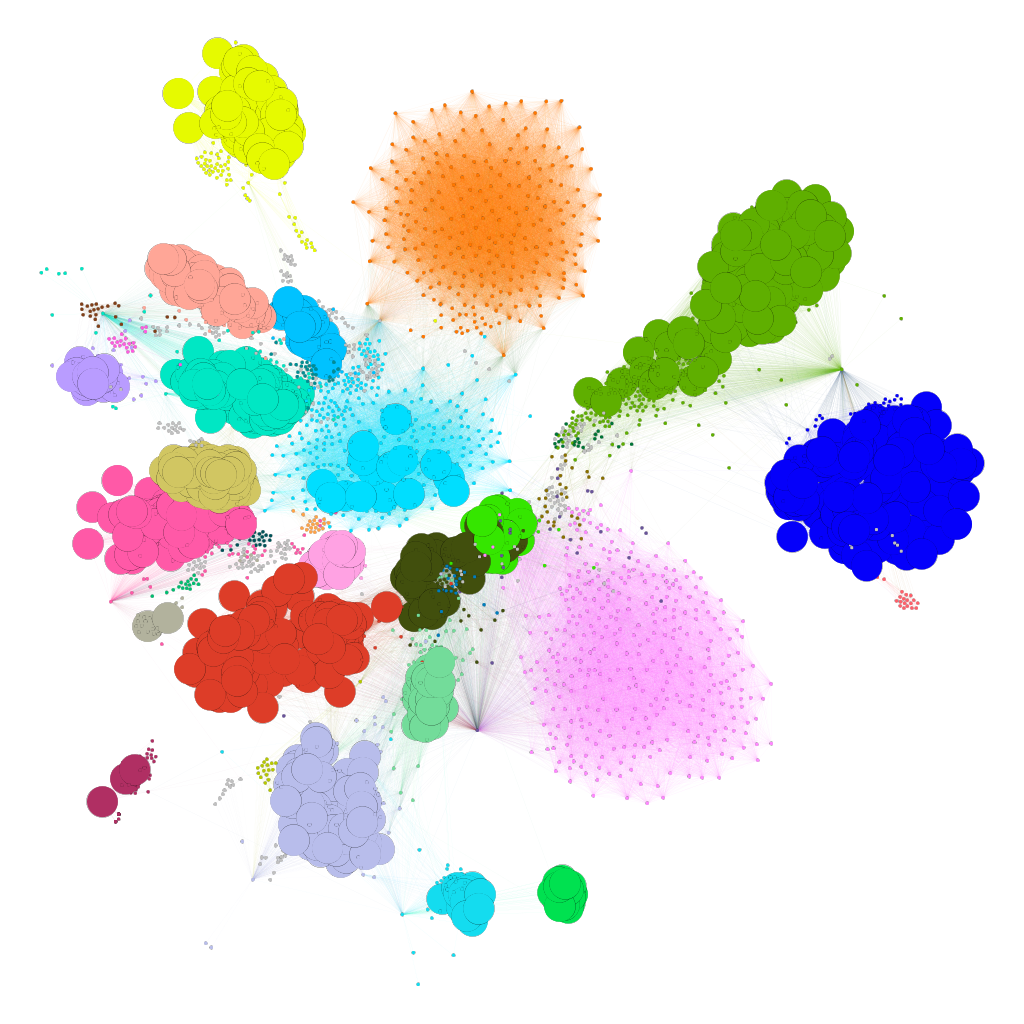}
    \caption{Nodes infected by Modularity Vitality at $f_o$ = 25\%}
    \label{fig:my_sub_4_EgoFacebook}
\end{subfigure}
\caption{Comparing the position of the nodes infected at low availability of resources, at $f_o$ = 1\% (top figures) and at high availability of resources, at $f_o$ = 25\% (bottom figures) in the Ego Facebook network. The bigger nodes in the left figure are the nodes picked by Community-based Mediator and the bigger nodes in right figure are the nodes picked by Modularity Vitality.}
\label{EgoFacebookVisualization}
\end{figure*}

\begin{figure*}[h!]
\centering
\begin{subfigure}{0.49\linewidth}
\centering
\includegraphics[height=6cm, width=6cm]{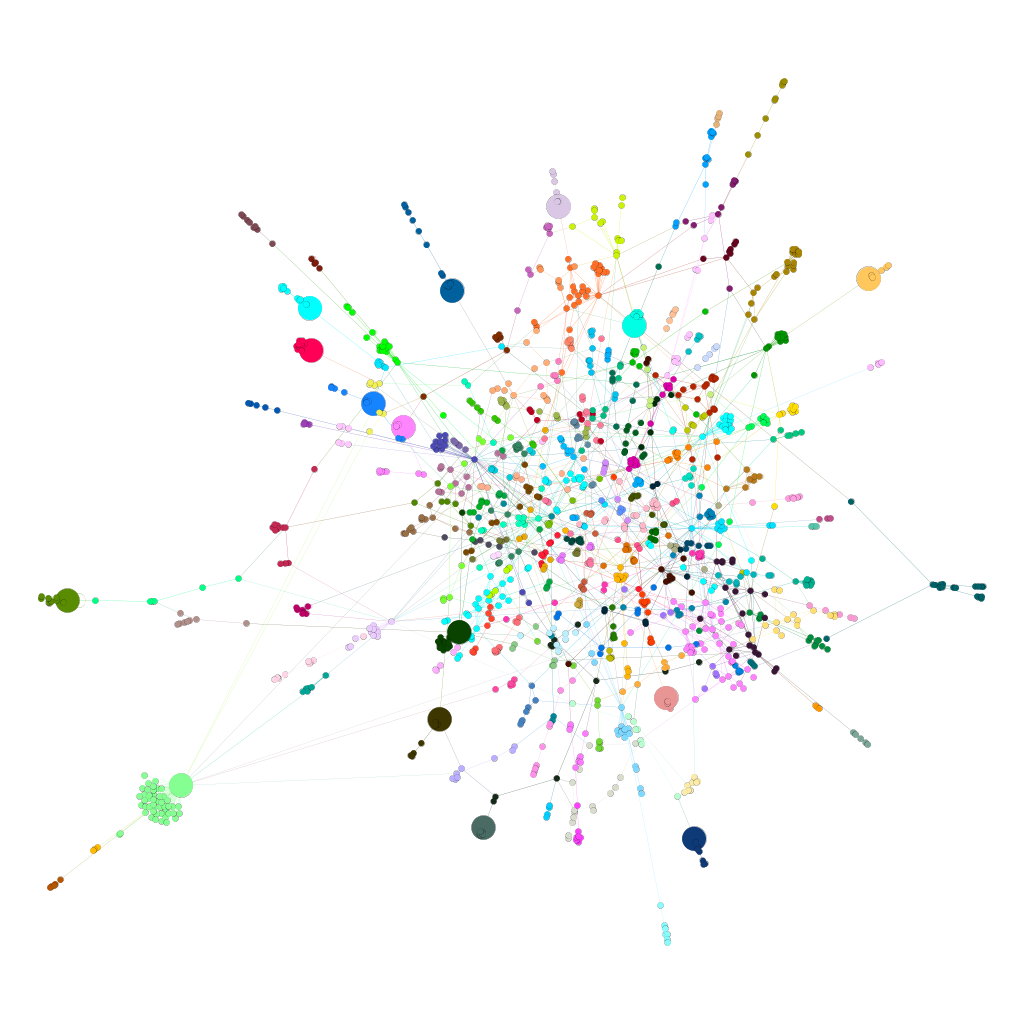}
    \caption{Nodes infected by Comm Centrality at $f_o$ = 1\%}
    \label{fig:my_sub_1_YeastProtein}
\end{subfigure}
\hfill
\begin{subfigure}{0.49\linewidth}
\centering
\includegraphics[height=6cm, width=6cm]{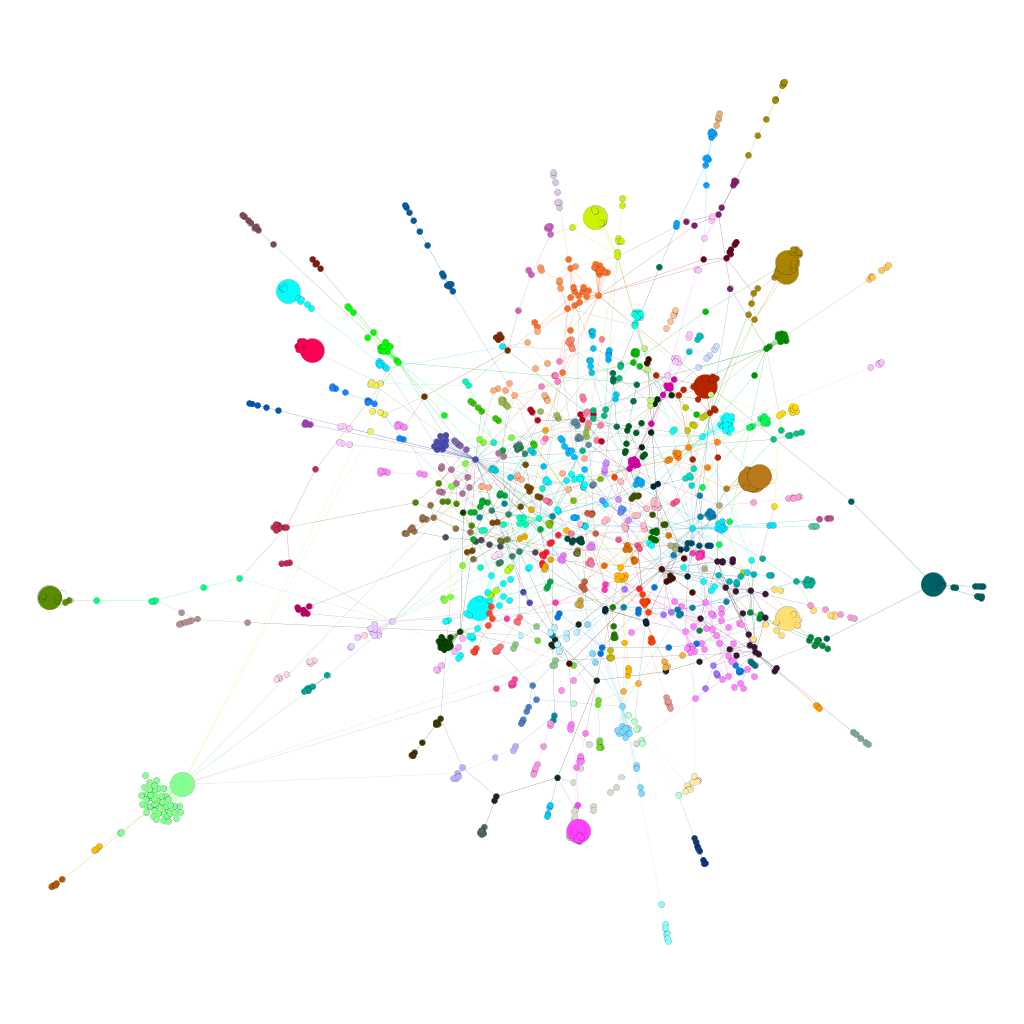}
    \caption{Nodes infected by Modularity Vitality at $f_o$ = 1\%}
    \label{fig:my_sub_2_YeastProtein}
\end{subfigure}
\vspace{0.5in} 
\begin{subfigure}{0.49\linewidth}
\centering
\includegraphics[height=6cm, width=6cm]{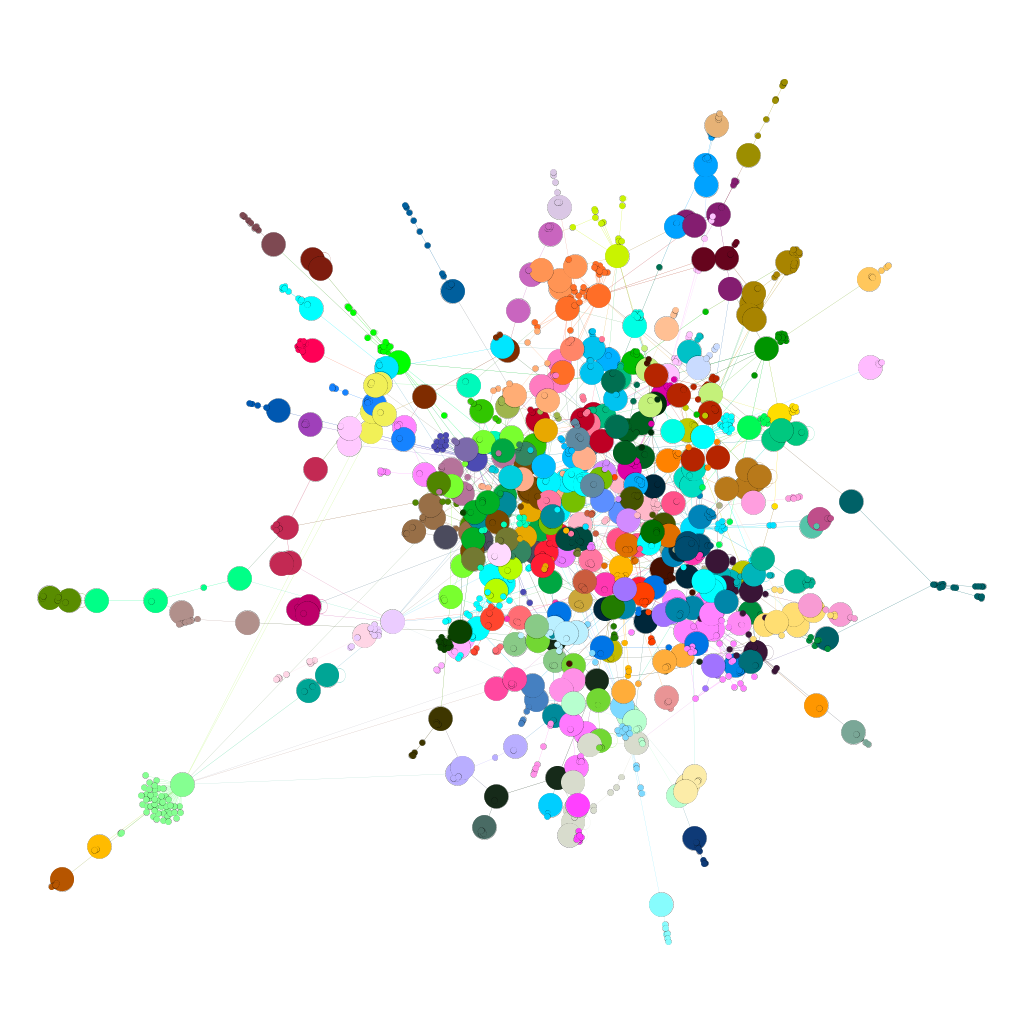}
    \caption{
    Nodes infected by Comm Centrality at $f_o$ = 25\%}
    \label{fig:my_sub_3_YeastProtein}
\end{subfigure}
\hfill
\begin{subfigure}{0.49\linewidth}
\centering
\includegraphics[height=6cm, width=6cm]{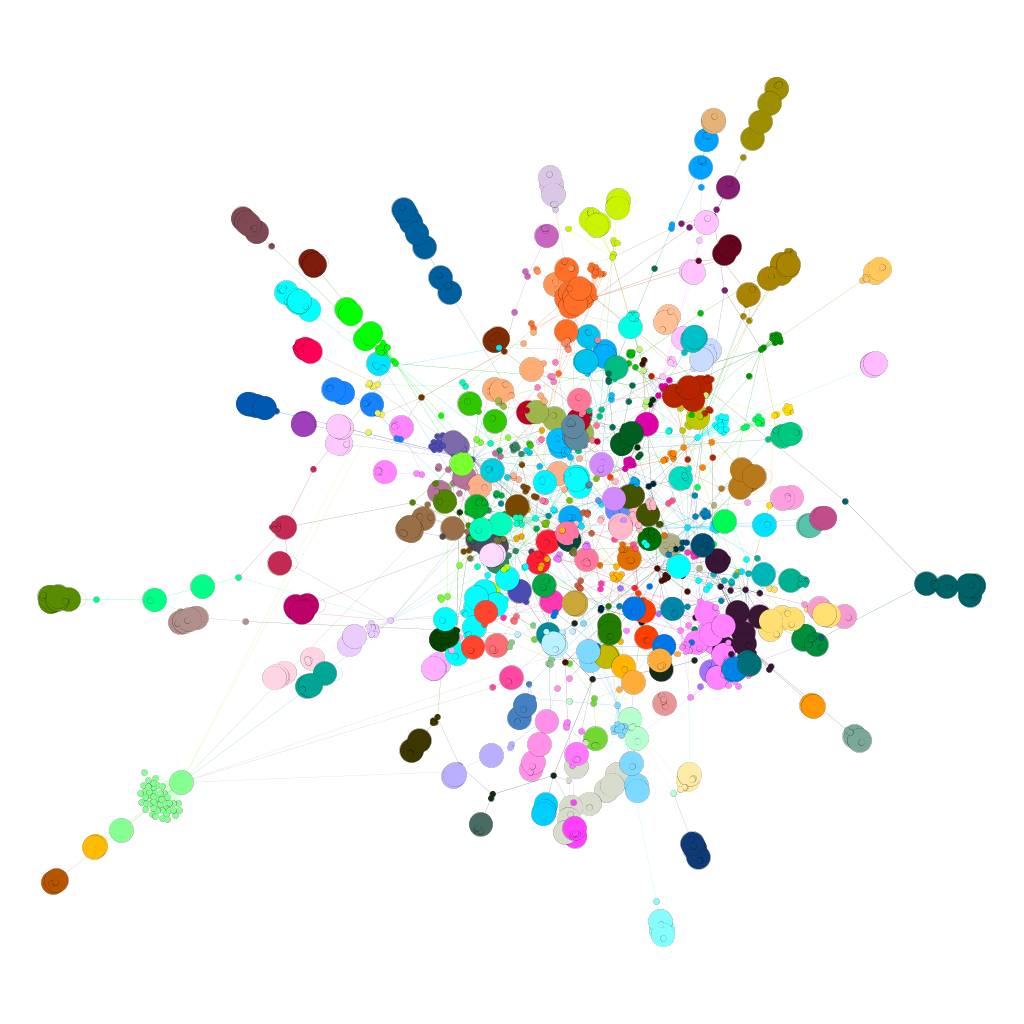}
    \caption{Nodes infected by Modularity Vitality at $f_o$ = 25\%}
    \label{fig:my_sub_4_YeastProtein}
\end{subfigure}
\caption{Comparing the position of the nodes infected at low availability of resources, at $f_o$ = 1\% (top figures) and at high availability of resources, at $f_o$ = 25\% (bottom figures) in the Yeast Protein network. The bigger nodes in the left figure are the nodes picked by Comm Centrality and the bigger nodes in right figure are the nodes picked by Modularity Vitality.}
\label{YeastProteinVisualization}
\end{figure*}

\begin{figure*}[h!]
\centering
\begin{subfigure}{0.49\linewidth}
\centering
\includegraphics[height=6cm, width=6cm]{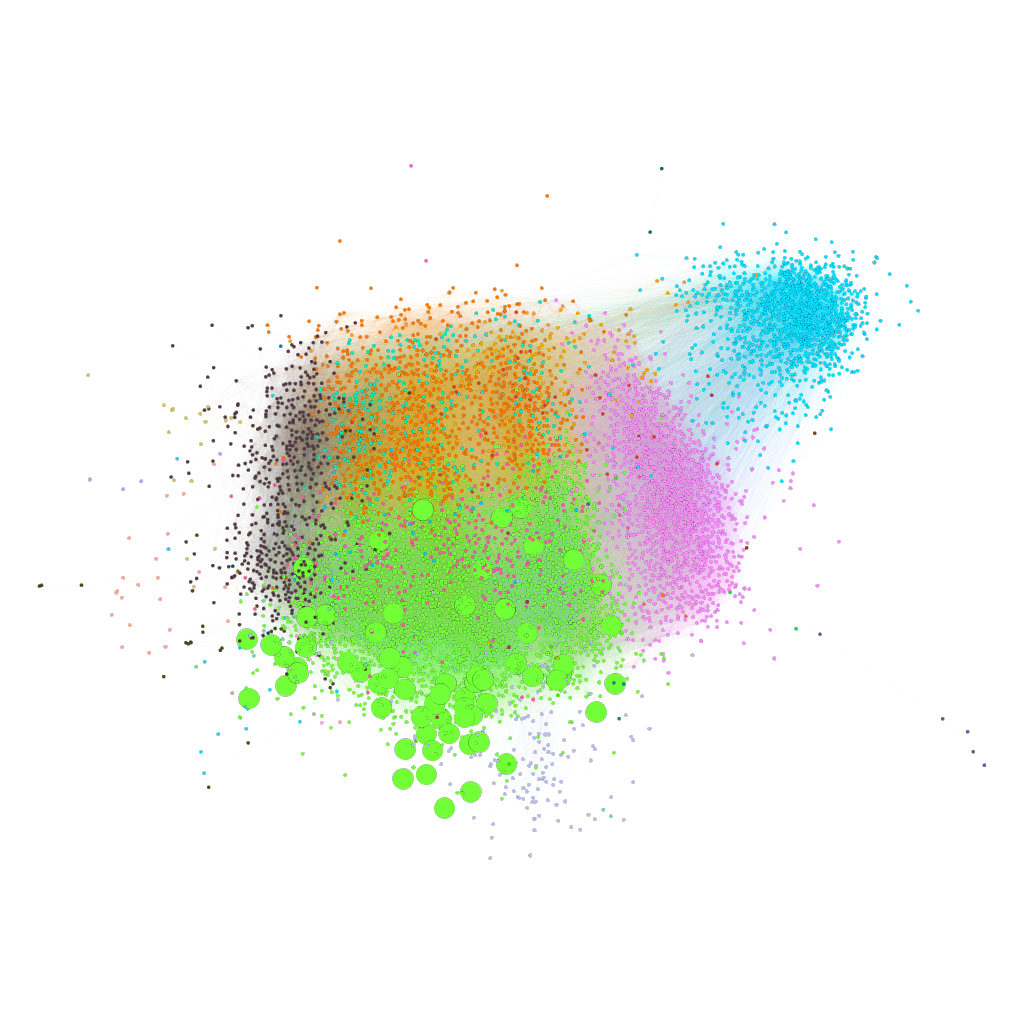}
    \caption{Nodes infected by Community Hub-Bridge at $f_o$ = 1\%}
    \label{fig:my_sub_1_Princeton}
\end{subfigure}
\hfill
\begin{subfigure}{0.49\linewidth}
\centering
\includegraphics[height=6cm, width=6cm]{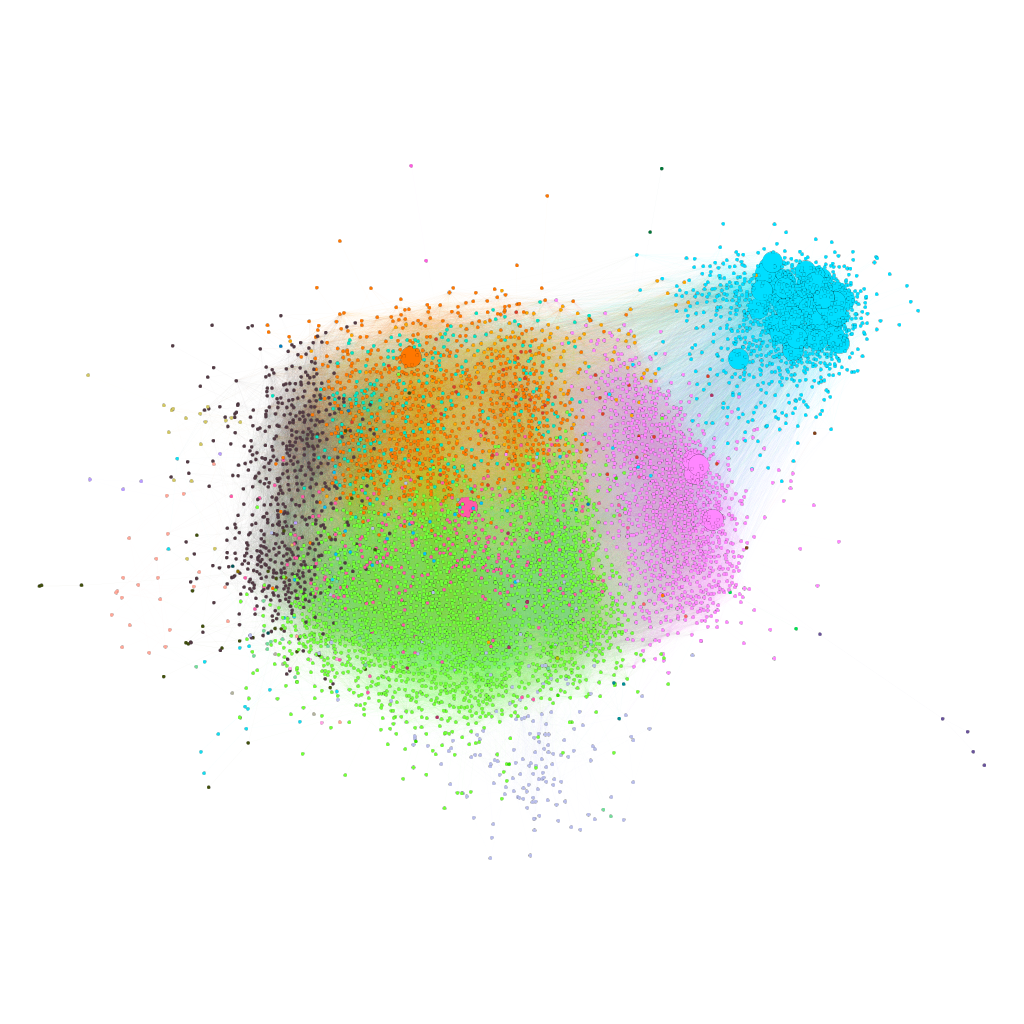}
    \caption{Nodes infected by Modularity Vitality at $f_o$ = 1\%}
    \label{fig:my_sub_2_Princeton}
\end{subfigure}
\vspace{0.5in} 
\begin{subfigure}{0.49\linewidth}
\centering
\includegraphics[height=6cm, width=6cm]{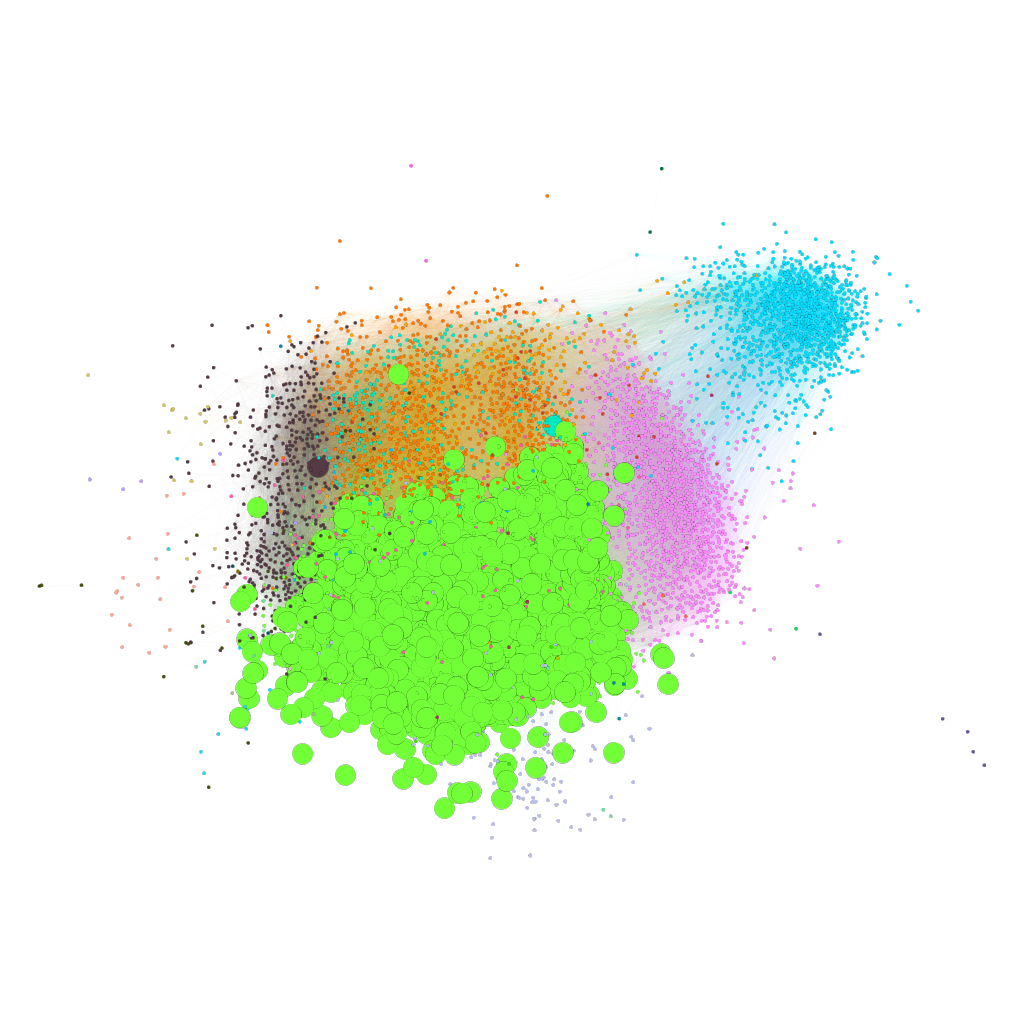}
    \caption{
    Nodes infected by Community-Hub Bridge at $f_o$ = 25\%}
    \label{fig:my_sub_3_Princeton}
\end{subfigure}
\hfill
\begin{subfigure}{0.49\linewidth}
\centering
\includegraphics[height=6cm, width=6cm]{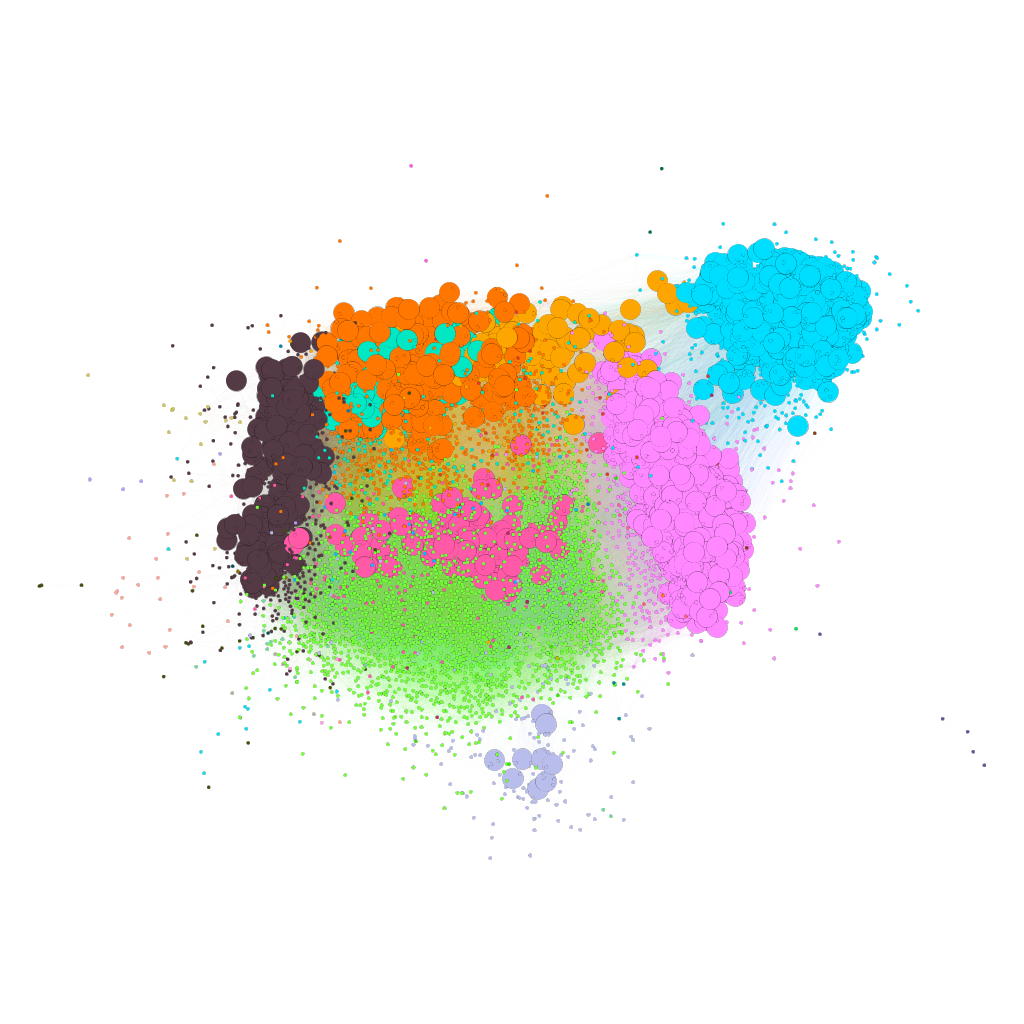}
    \caption{Nodes infected by Modularity Vitality at $f_o$ = 25\%}
    \label{fig:my_sub_4_Princeton}
\end{subfigure}
\caption{Comparing the position of the nodes infected at low availability of resources, at $f_o$ = 1\% (top figures) and at high availability of resources, at $f_o$ = 25\% (bottom figures) in the Princeton network. The bigger nodes in the left figure are the nodes picked by Community Hub-Bridge and the bigger nodes in right figure are the nodes picked by Modularity Vitality.}
\label{PrincetonVisualization}
\end{figure*}

\begin{figure*}[h!]
\centering
\begin{subfigure}{0.49\linewidth}
\centering
\includegraphics[height=6cm, width=6cm]{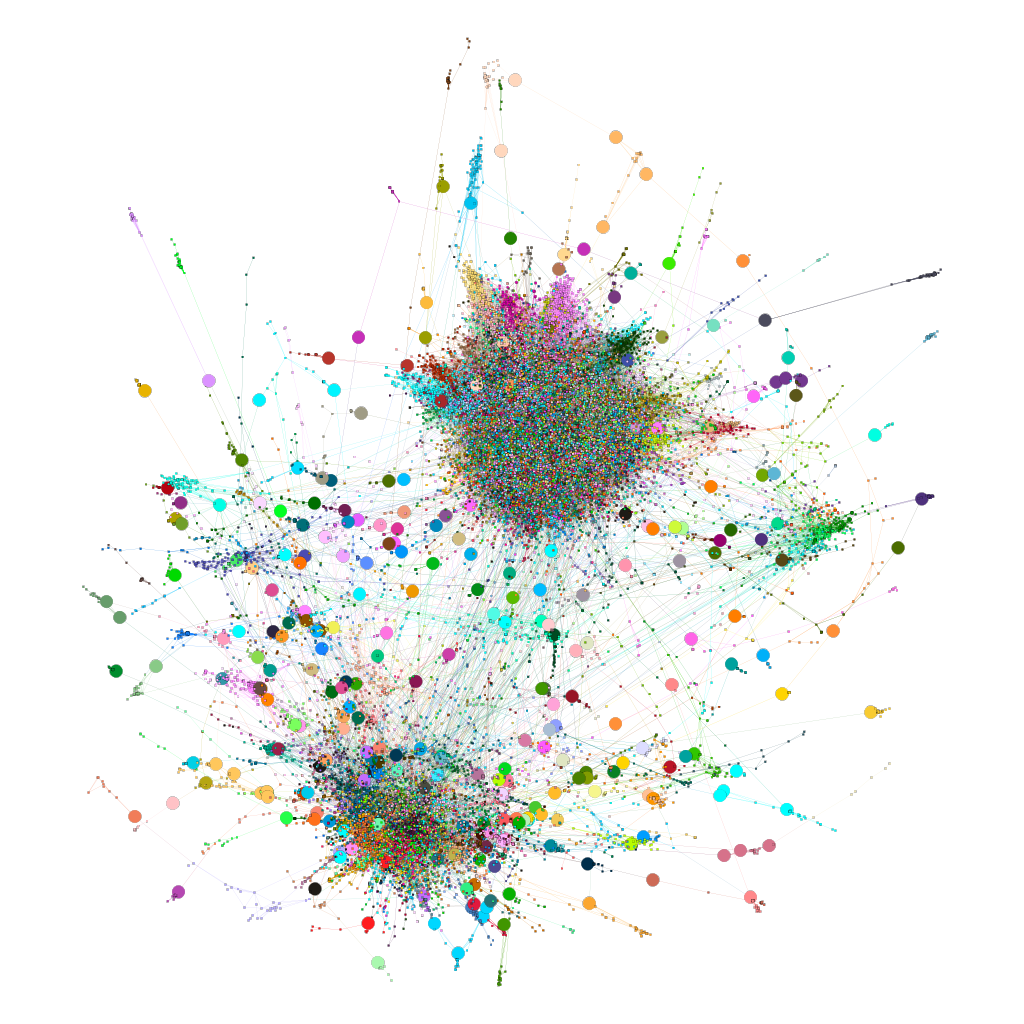}
    \caption{Nodes infected by Comm Centrality at $f_o$ = 1\%}
    \label{fig:my_sub_1_DeezerEU}
\end{subfigure}
\hfill
\begin{subfigure}{0.49\linewidth}
\centering
\includegraphics[height=6cm, width=6cm]{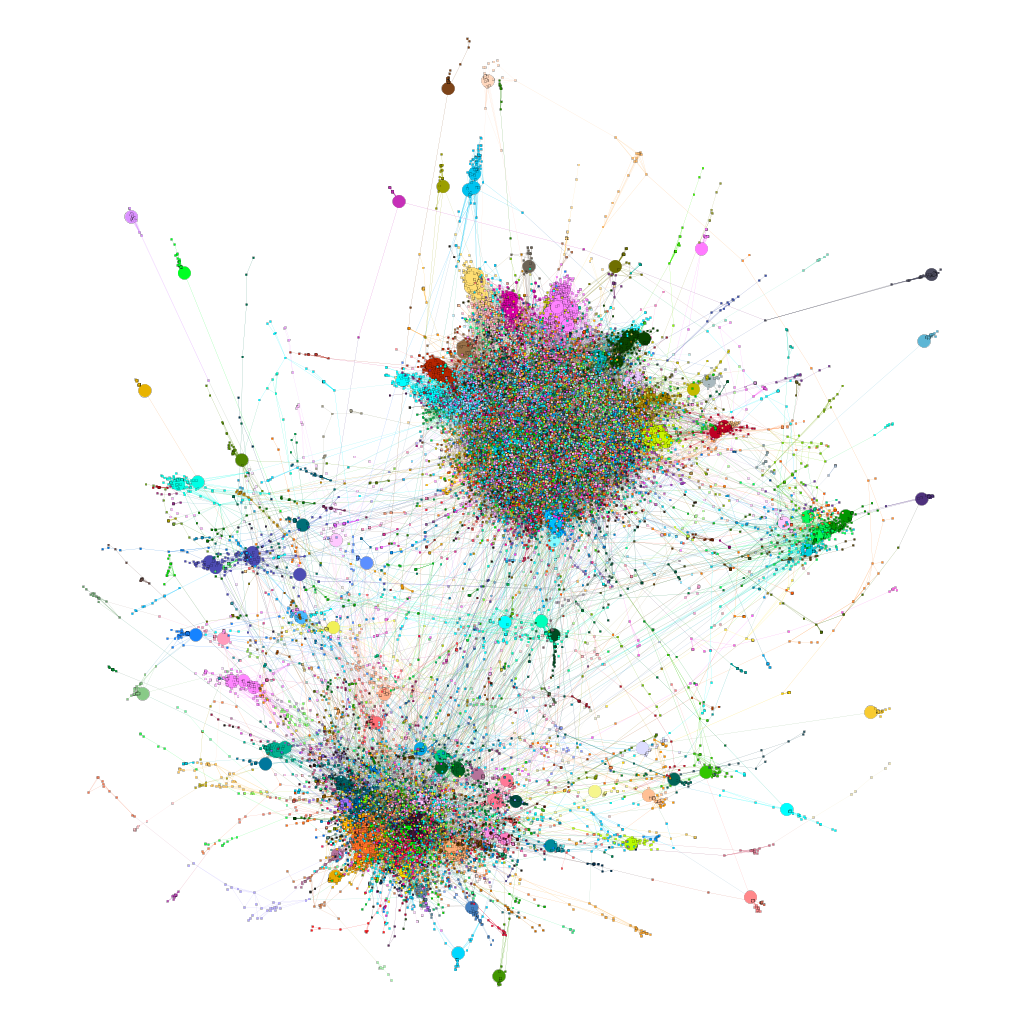}
    \caption{Nodes infected by Modularity Vitality at $f_o$ = 1\%}
    \label{fig:my_sub_2_DeezerEU}
\end{subfigure}
\vspace{0.5in} 
\begin{subfigure}{0.49\linewidth}
\centering
\includegraphics[height=6cm, width=6cm]{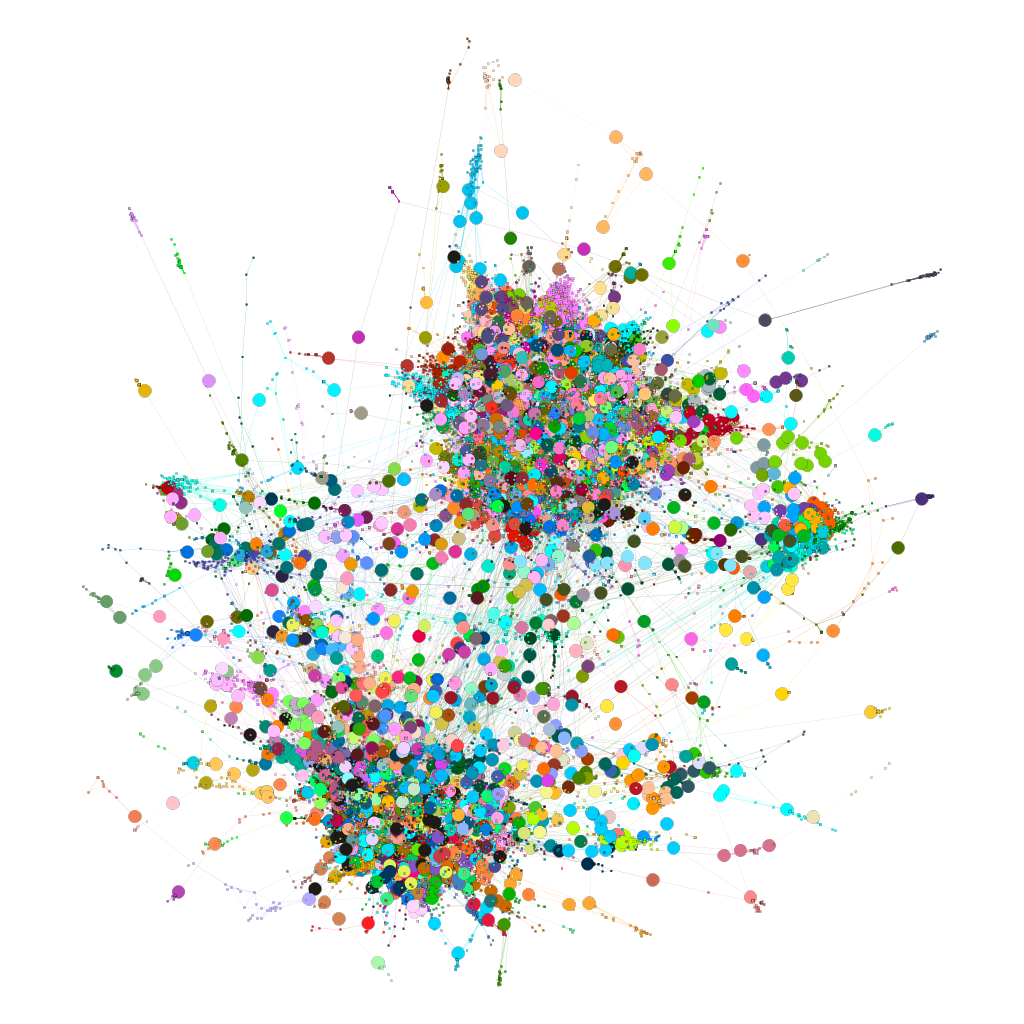}
    \caption{
    Nodes infected by Comm Centrality at $f_o$ = 25\%}
    \label{fig:my_sub_3_DeezerEU}
\end{subfigure}
\hfill
\begin{subfigure}{0.49\linewidth}
\centering
\includegraphics[height=6cm, width=6cm]{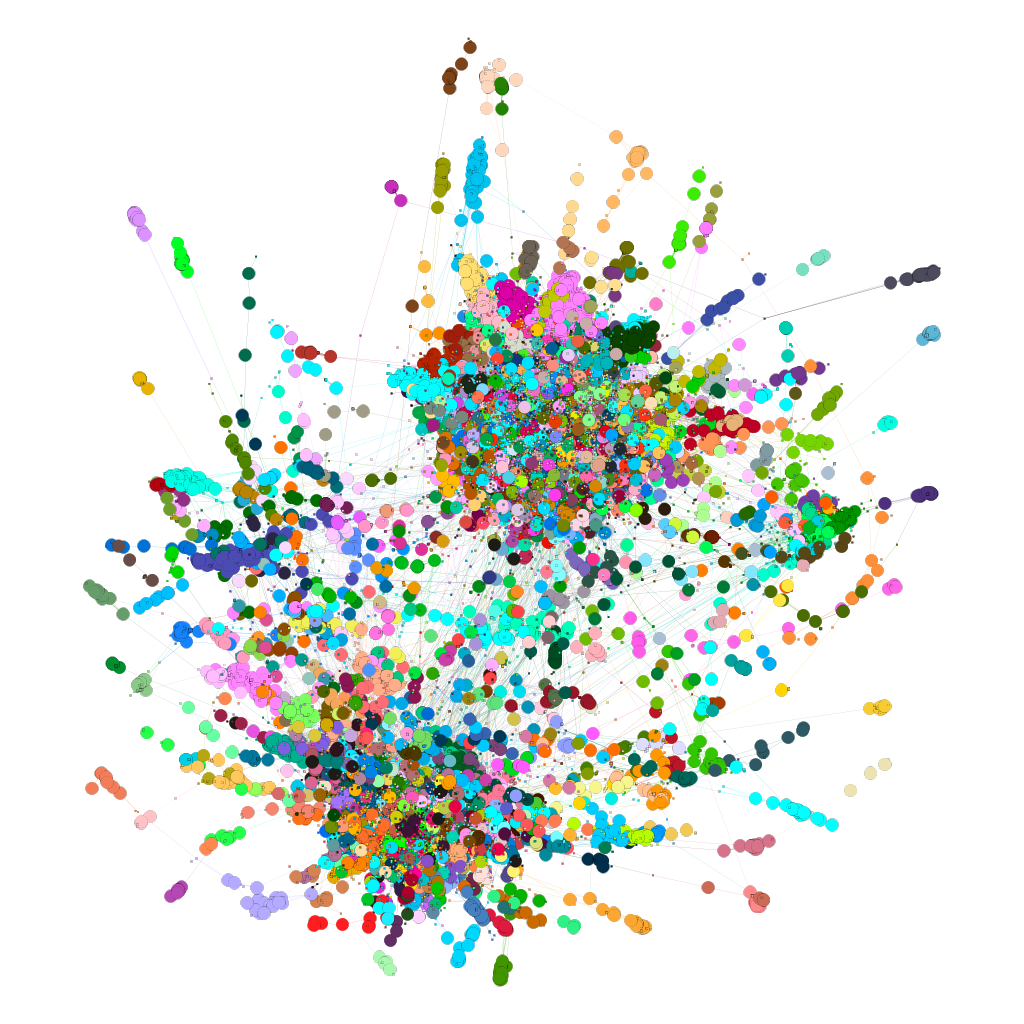}
    \caption{Nodes infected by Modularity Vitality at $f_o$ = 25\%}
    \label{fig:my_sub_DeezerEU}
\end{subfigure}
\caption{Comparing the position of the nodes infected at low availability of resources, at $f_o$ = 1\% (top figures) and at high availability of resources, at $f_o$ = 25\% (bottom figures) in the DeezerEU network. The bigger nodes in the left figure are the nodes picked by Comm Centrality and the bigger nodes in right figure are the nodes picked by Modularity Vitality.}
\label{DeezerEUVisualization}
\end{figure*}

\begin{figure*}[h!]
\centering
\begin{subfigure}{0.49\linewidth}
\centering
\includegraphics[height=6cm, width=6cm]{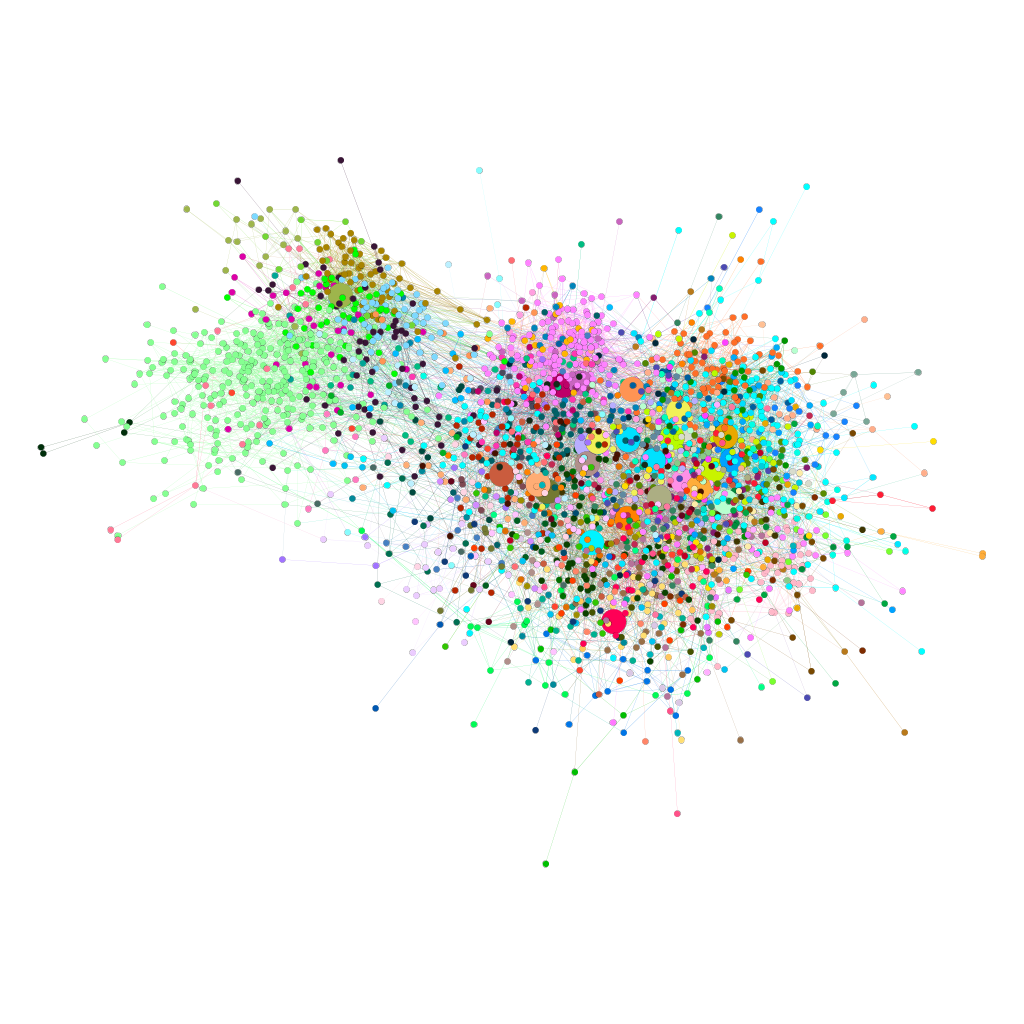}
    \caption{Nodes infected by Participation Coefficient at $f_o$ = 1\%}
    \label{fig:my_sub_1_Adol}
\end{subfigure}
\hfill
\begin{subfigure}{0.49\linewidth}
\centering
\includegraphics[height=6cm, width=6cm]{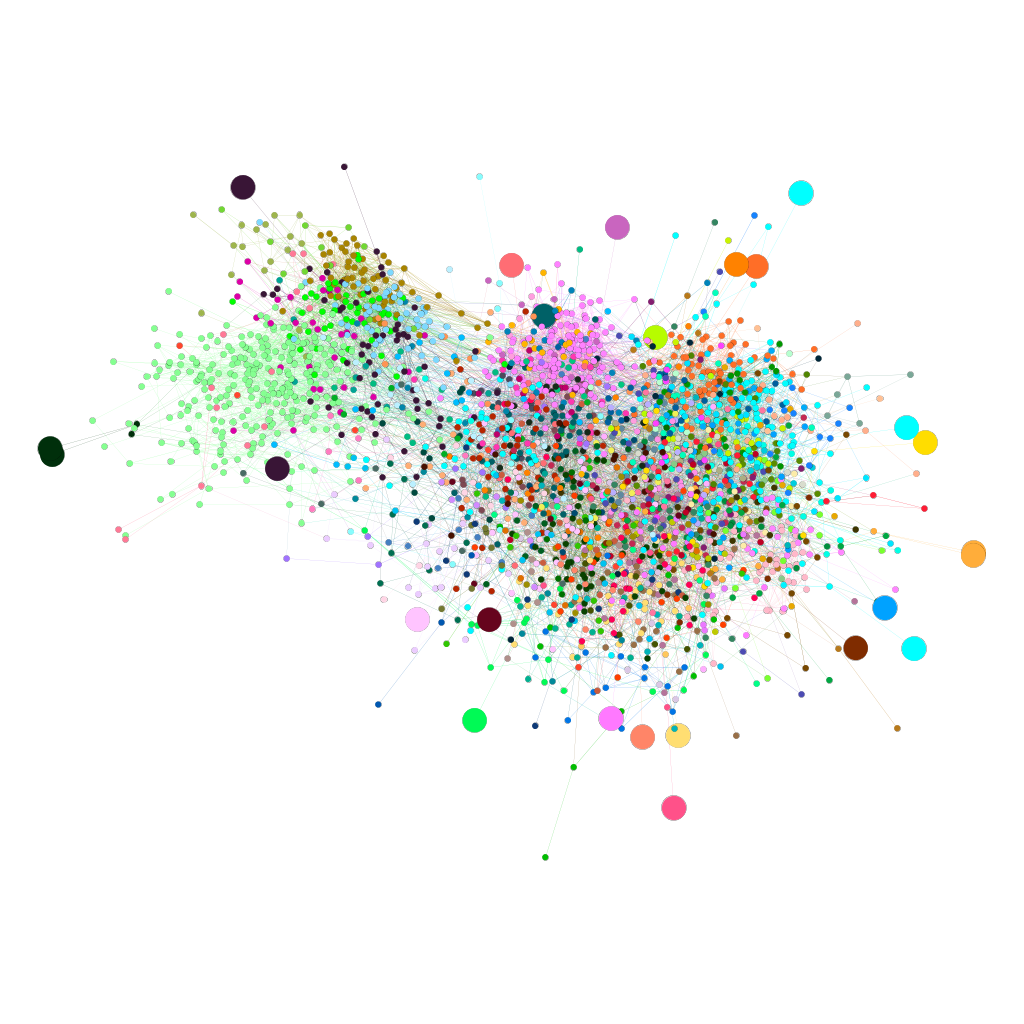}
    \caption{Nodes infected by Modularity Vitality at $f_o$ = 1\%}
    \label{fig:my_sub_2_Adol}
\end{subfigure}
\vspace{0.5in} 
\begin{subfigure}{0.49\linewidth}
\centering
\includegraphics[height=6cm, width=6cm]{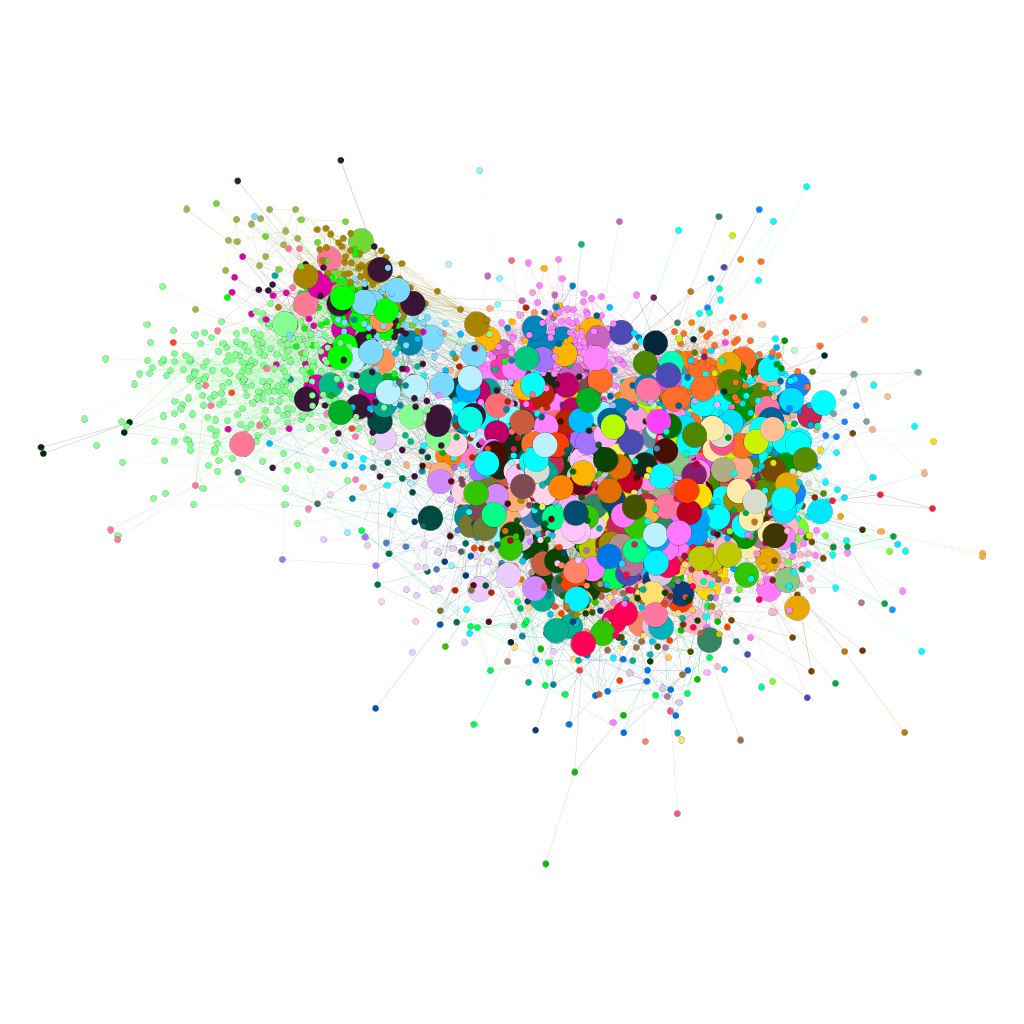}
    \caption{
    Nodes infected by Participation Coefficient at $f_o$ = 25\%}
    \label{fig:my_sub_3_Adol}
\end{subfigure}
\hfill
\begin{subfigure}{0.49\linewidth}
\centering
\includegraphics[height=6cm, width=6cm]{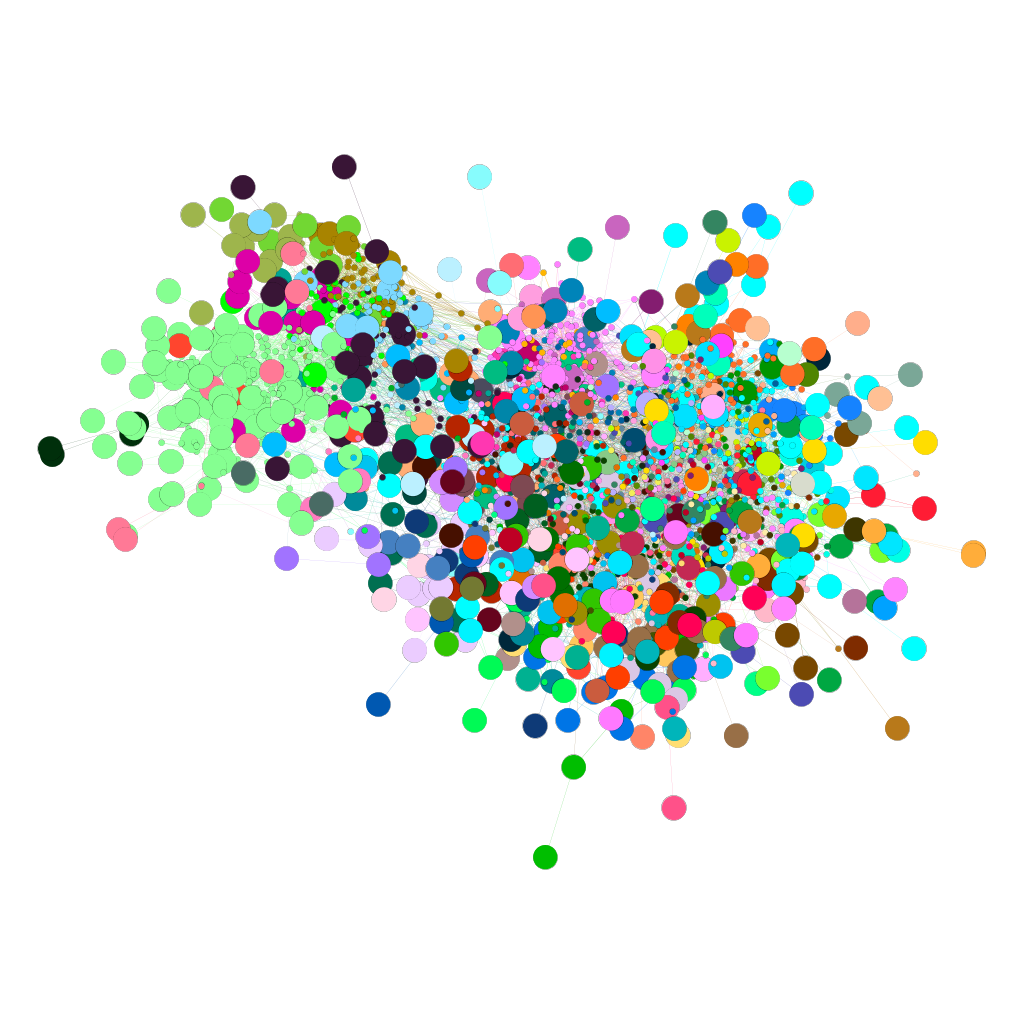}
    \caption{Nodes infected by Modularity Vitality at $f_o$ = 25\%}
    \label{fig:my_sub_4_Adol}
\end{subfigure}
\caption{Comparing the position of the nodes infected at low availability of resources, at $f_o$ = 1\% (top figures) and at high availability of resources, at $f_o$ = 25\% (bottom figures) in the Adolescent Health network. The bigger nodes in the left figure are the nodes picked by Participation Coefficient and the bigger nodes in right figure are the nodes picked by Modularity Vitality.}
\label{AdolVisualization}
\end{figure*}

To investigate the variation across networks of the correlation pattern between community-aware centrality measures, we calculate the Pearson correlation of the networks' heatmaps. It appears that networks with a strong community structure do not correlate well with the others. In addition, the ones correlating the most with the others have a medium or weak community structure. A linear regression analysis corroborates these findings.

We show that the community structure strength is a crucial feature driving the correlation variation between community-aware centrality measures. In networks with a strong community structure, community-aware centrality measures do not correlate well. 
Correlation between the centrality measures increases when the community structure gets weaker. Indeed, the weaker the community structure, the fewer differences between the various community-aware centrality measures. This result is quite intuitive. For example, consider Comm Centrality and Community-based Mediator. The first favors bridges against hubs by raising the inter-community links proportion to the power of 2. The second exploits the entropy of the intra-community and inter-community connections. If a network has a well-defined community structure, these two measures select different ``bridges." Indeed, in Comm Centrality, intra-community links might not play a significant role in Community-based Mediator, which is affected by a simple change in the latter. Now consider a network with a weak community structure. In this case, a node's local and global influence are less diverse since many nodes from different communities are connected. Consequently, the Comm Centrality and Community-based Mediator extract on average more similar information than in a network with a strong community structure. 
Note that these results are in the same vein that findings reported in previous works \cite{rajeh2021characterizing, rajeh2021correlated}. Extensive investigation has shown that the correlation variation between classical and community-aware centrality measures is mainly related to the community structure strength.

We compare the diffusion ability of the various community-aware centrality measures using SIR simulations. Results show that Modularity Vitality obtains the most significant outbreak difference with the baseline in networks with strong, medium, and strong community structure. However, it may take not take the lead when a small fraction of nodes is initially infected. Note that Modularity Vitality used in this study targets hubs. The low correlation values observed with the other community-aware centrality measures suggest that they mainly focus on bridges.

To further compare the distribution of the top nodes extracted by Modularity Vitality
to its alternatives, we visualize the six representative networks of Fig. \ref{Fig2-SIR} with their communities reported in different colors. The big nodes represent the fraction of top nodes initially infected based on their respective community-aware centrality measure.

Figures \ref{EUAirlinesVisualization} and \ref{EgoFacebookVisualization}, represent respectively EU Airlines and Ego Facebook. These networks have a strong community structure. In EU Airlines, Modularity Vitality outperforms the other measures with a small fraction of initially infected nodes ($f_o$ = 1\%). One can see in Fig. \ref{EUAirlinesVisualization} representing the EU Airlines network that  Modularity Vitality neglects hubs at the core of the largest community (purple). Instead, it targets hubs in all the other small communities. This remark still holds when the fraction of initially infected nodes grows ($f_o$ = 25\%). Indeed, the few nodes selected in the purple community are in its periphery in this situation. Modularity Vitality does not favor hubs in the purple community since their removal does not severely impact the modularity of the network. In other words, since the purple community is very dense, removing a hub from another community has a higher effect on modularity.

In contrast, the left figure shows nodes chosen by K-shell with Community. One can see that, whatever the value of the fraction of initially infected nodes, all the extracted nodes are at the core of the purple community. Consequently, K-shell with Community exhibits poor performance. Indeed, the epidemic does not reach the other communities and the nodes located at the network's periphery. Overall, the more scattered the top centrality nodes, the higher the probability that the infection reaches the various part of the network.

Although the Ego Facebook network is also a network with a strong community structure, its topology is quite different than EU Airlines. Indeed, the size of the communities is more balanced, while in EU Airlines, it is more nonhomogeneous. In this situation, the hubs chosen by Modularity Vitality, even though they are from different communities, don't result in the highest epidemic outbreak. Indeed, there is no single winner in Ego Facebook. First, Community-based Mediator leads to the highest epidemic outbreak. It is followed by Comm Centrality and by Participation Coefficient. Consider the upper subfigures of Fig. \ref{EgoFacebookVisualization} where 1\% of the nodes are infected. One can see that Community-based Mediator selects nodes located at the borders of the communities.
In contrast, Modularity Vitality chooses hubs in small communities at the expense of those in large communities.  The epidemic spreading is more effective with Community-based Mediator in this case. The same behavior persists with more initially infected nodes  (bottom subfigures). Modularity Vitality neglects the pink and orange communities, while Community-based Mediator infects all the communities. Overall, Modularity Vitality favors hubs in smaller communities since their removal has a higher impact on the network's modularity. Consequently, as it neglects larger communities with numerous hubs, an infection may not reach them.

Figures \ref{YeastProteinVisualization} and \ref{PrincetonVisualization} represent Yeast Protein and Princeton networks, respectively. These networks have a medium community structure strength. In Yeast Protein, Comm Centrality performs slightly better than Modularity Vitality with a low proportion of initially infected nodes ($f_o$ = 1\%).  As this proportion increases, Modularity Vitality takes the lead. One can see that Modularity Vitality covers more communities at $f_o$ = 25\%. Indeed, Modularity Vitality infects nodes located in the network periphery in numerous small communities. In return, the epidemic outbreak reaches parts of the network that are not easily reachable from the core communities. Concerning the Princeton network, Community Hub-Bridge has a slightly higher epidemic outbreak than Modularity Vitality and Participation Coefficient till $f_o$ = 33\%. After that, Modularity Vitality takes over. We can see that Community Hub-Bridge selects nodes in the green community, while Modularity Vitality picks most nodes in the blue community. Nevertheless, even with a small fraction of infected nodes, it covers more communities. In the end, Modularity Vitality leads to a higher epidemic outbreak as the fraction of initially infected nodes grows. One can see that initially infected nodes surround the neglected green community. Thus, infections will most likely enter it through different nodes.

The limiting case of a network with a weak community structure is a network with no community structure. In this situation, the influence of the community structure is limited. The best strategy to reach the various zones of the network is to target the most distant nodes. Indeed, this is what Modularity Vitality shows in the networks DeezerEU and Adolescent Health, illustrated in Figures \ref{DeezerEUVisualization} and \ref{AdolVisualization}, respectively. In DeezerEU, at a small infection rate ($f_o$ = 1\%), the nodes selected by Come Centrality are more distant compared to Modularity Vitality. It results in a better performance. At $f_o$ = 25\%, nodes chosen by Modularity Vitality are more evenly distributed across the network, resulting in higher performance. One observes a similar behavior in the Adolescent Health network illustrated in Fig. \ref{AdolVisualization}. In this case, Modularity Vitality outperforms the other measures, whatever the proportion of initially infected nodes.

To summarize, Modularity Vitality achieves the highest epidemic outbreak when there is a fair proportion of initially infected nodes. If there are not enough resources, it tends to underestimate large communities densely connected. Its performance is more pronounced as the community structure gets weaker. It is due to its natural behavior of selecting hubs that are distant from each other. It tends to pick hubs in communities containing few hubs since their removal significantly affects the network's modularity. Nonetheless, targeting a small number of hubs is not enough below a specific fraction of infections (or resources). Here, it is better to target bridges, especially if the network has a strong or medium community structure strength. Indeed, there are enough external connections between communities in these networks to diffuse in all the communities of the network. Hence, infected bridges spread the epidemic to communities more effectively than hubs.

It is worth noticing that a couple of community-aware centrality measures don't perform better than the baseline. It is evident with K-shell with Community in Ego Facebook, U.S. Airports, Facebook Politician Pages, Facebook Friends, U.S. Power Grid, Yeast Protein, Facebook Organizations, and DeezerEU. To a lesser extent, it applies to Community Hub-Bridge and Community-based Centrality. Sometimes, community-aware centrality measures perform poorly, even though they incorporate relevant information about the community structure.  For example, in K-shell with Community, many nodes share the same $k$-shell (or $k$-core) value. This behavior is illustrated by K-shell with Community in the EU Airlines network in Fig. \ref{EUAirlinesVisualization}. Therefore, infecting one or the other on the same level may not lead to a higher epidemic spreading than infecting more distant nodes (i.e., not on the same level). Similarly is the case for Community-based Centrality, which shrinks the inter-community and intra-community links based on their community sizes. So, many nodes receive very similar centrality values. Moreover, these nodes may be close to each other. Note that this is only true when one considers a multiple-spreader problem where several nodes are infected simultaneously. Nonetheless, if the infection starts from a single node, nodes that are the most embedded in the network may result in the highest epidemic outbreak \cite{rajeh2021comparing, kitsak2010identification}. For more insights on the single-spreader problem versus the multiple-spreader problem, one can consult the following literature \cite{everett1999centrality, everett2005extending}.

\begin{figure*}[h]
\centerline{\includegraphics[width=1\linewidth, height=2 in]{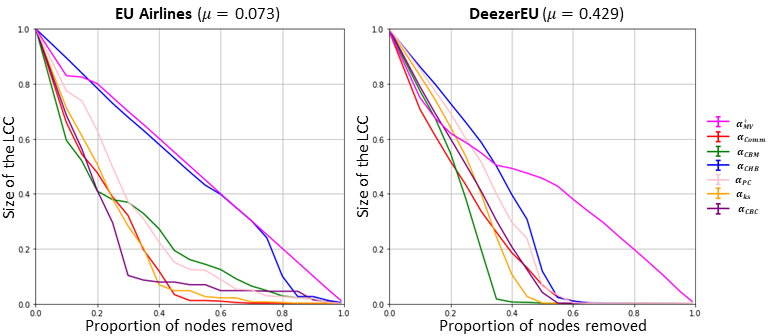}}
\caption{The size of largest connected component (LCC) of the EU Airlines and DeezerEU networks. The community-aware centrality measures are:  Modularity Vitality targeting hubs = $\alpha^+_{MV}$, Comm Centrality = $\alpha_{Comm}$, Community‑based Mediator = $\alpha_{CBM}$, Community Hub-Bridge = $\alpha_{CHB}$, Participation Coefficient = $\alpha_{PC}$, K-shell with Community = $\alpha_{ks}$,  and Community-based Centrality = $\alpha_{CBC}$.}
\label{LCCFig}
\end{figure*}

Finally, we compute the evolution of the size of the largest connected component (LCC) when removing a given fraction of top-ranked nodes using the various community-aware centrality measures. Figure \ref{LCCFig} illustrates the general trend observed in two networks (EU Airlines and DeezerEU). According to the LCC, Modularity Vitality performs poorly in dismantling the network in both cases. In contrast, considering a diffusion process based on the SIR model (Fig. \ref{Fig2-SIR}), Modularity Vitality results in the highest epidemic outbreak once a specific fraction of initially infected nodes is reached. One should be cautious when looking at these results, which may seem contradictory. Remember that the LCC is a conservative upper bound corresponding to the worst-case scenario with no information about the diffusion process. In this situation, strategies targeting bridges such as Community-based Mediator or nodes embedded in the large communities using Community-based Centrality are more suitable strategies for efficient network dismantling.


Remember that Modularity Vitality targets hubs in small communities in priority because they induce the highest variation of modularity. The largest communities are affected in the SIR diffusion process when the fraction of initially infected nodes is high enough. However, the size of the LCC is barely affected compared to the other measures prioritizing bridges. Consequently, as the most prominent communities are not involved in the early stage of the infection process, the size of the LCC remains significant even if the epidemic diffusion is contained. Note that if Modularity Vitality uses the bridges-first ranking scheme, networks may disintegrate much quicker \cite{modvitality}. These results demonstrate that one should be cautious in using the LCC to compare the performance of centrality measures. If the diffusion process is known, one should use it for a more realistic investigation.

\section{Conclusion}
\label{sec:Conc}

Identifying influential nodes in a network has always attracted scientists due to its broad theoretical and applicative dimensions. Several previous studies investigate the relations between classical centrality measures. In contrast, this work focuses on the interplay between popular community-aware centrality measures using real-world networks from various domains and diverse topological characteristics.

Results show that the correlation between the community-aware centrality measures ranges from negative to positive. Modularity Vitality exhibits low negative correlation values with the other community-aware centrality measures. It differs from the others mainly because it is a signed centrality. Furthermore, it explicitly targets hubs. The other community-aware centrality measures target bridges or hubs and bridges simultaneously. Their correlation is positive. It generally ranges from weak positive to medium positive values. The community structure strength affects the correlation pattern. The stronger the community structure strength, the lower the correlation. Conversely, networks with weak community structures exhibit higher correlation.

If enough resources are available, SIR simulations suggest targeting hubs using Modularity Vitality instead of targeting bridges.  This strategy is more effective since it extracts from different distant communities. It is even more effective in networks with a weak community structure. Indeed, rather than selecting nodes in the network core, it allows covering the various modules of the network and leads to a higher epidemic outbreak. In networks with strong to medium community structures, the choice of a community-aware centrality measure is more dependent on the resources at hand. It is better to use Comm Centrality that targets bridges rather than hubs when resources are limited. Indeed, in these situations, the epidemic spreads into the communities through inter-community links between communities.

These findings have multiple implications. First, one can choose the adequate measure knowing the network’s community structure strength and the availability resources at hand. Second, they open new perspectives in the design of community-aware centrality measures adapted to specific networks' mesoscopic characteristics. Third, they pave the way for future investigations on the interplay between various diffusion models and the networks' mesoscopic topological properties.

\section{Declarations}
\textbf{Funding}  Not applicable.

\textbf{Conflicts of interest/Competing interests} Author HC has served on the editorial board of Quality and Quantity. The authors declare that they have no competing interests. 

\textbf{Availability of data and material} The datasets used in this article are all freely accessible in the cited resources.

\textbf{Code availability} The code of the following study is accessible via GitHub: https://github.com/StephanyRajeh/CompAnalysisCACM

\textbf{Authors' contributions} All authors have contributed to this work. SR implemented the models and the analyses. All authors participated in the formulation and writing of this paper. All authors approved the final manuscript.

\textbf{Ethics approval} Not applicable.

\textbf{Consent to participate} Not applicable.

\textbf{Consent for publication} Not applicable.

\bibliographystyle{spmpsci}      

\end{document}